\documentclass[reprint, superscriptaddress, preprintnumbers, aps, prl]{revtex4-1}
\pdfoutput=1

\usepackage[utf8]{inputenc}
\usepackage{setup}

\DeclareRobustCommand{\ptv}{{\ensuremath{p_\rT^V}}\xspace}
\DeclareRobustCommand{\ptjo}{{\ensuremath{p_\rT^{j1}}}\xspace}
\DeclareRobustCommand{\ptj}{{\ensuremath{p_\rT^j}}\xspace}

\DeclareRobustCommand{\pt}{{\ensuremath{p_\rT}}\xspace}

\DeclareRobustCommand{\NNLOJET}{NNLO\scalebox{.8}{JET}\xspace}

\begin{document}

\author{A.\ Gehrmann--De Ridder}%
  \affiliation{Institute for Theoretical Physics, ETH, CH-8093 Z\"urich, Switzerland}%
  \affiliation{Department of Physics, University of Z\"urich, CH-8057 Z\"urich, Switzerland}%
\author{T.\ Gehrmann}%
  \affiliation{Department of Physics, University of Z\"urich, CH-8057 Z\"urich, Switzerland}%
\author{E.\ W.\ N.\ Glover}%
  \affiliation{Institute for Particle Physics Phenomenology, Durham University, Durham DH1 3LE, UK}%
\author{A.\ Huss}%
  \affiliation{Theoretical Physics Department, CERN, CH-1211 Geneva 23, Switzerland}%
\author{D.\ M.\ Walker}%
  \affiliation{Institute for Particle Physics Phenomenology, Durham University, Durham DH1 3LE, UK}%

\preprint{IPPP/19/2, ZU-TH 03/19, CERN-TH-2019-004}

\title{Vector Boson Production in Association with a Jet at Forward Rapidities}

\begin{abstract}
  Final states with a vector boson and a hadronic jet allow one to infer the Born-level kinematics of
  the underlying hard scattering process, thereby probing the partonic structure of the colliding protons. At forward
  rapidities, the parton collisions are highly asymmetric and resolve the parton distributions at very large
  or very small momentum fractions, where they are less well constrained by other processes.
  Using theory predictions accurate to
  next-to-next-to-leading order (NNLO) in QCD for both \PWpm and \PZ production in association with a jet at large rapidities at the LHC, we perform a detailed phenomenological analysis of recent LHC measurements. The increased
  theory precision allows us to clearly identify
  specific kinematical regions where the description of the data is insufficient.
  By constructing
  ratios and asymmetries of these cross sections, we aim to identify possible origins of the deviations, and highlight
  the potential impact of the data on improved determinations of parton distributions.
\end{abstract}

\maketitle

\section{Introduction}

The production of a vector boson in association with a hadronic jet is the simplest hadron-collider process
that probes both the strong and electroweak interactions at Born level.
It has been measured extensively at the
Tevatron~\cite{Aaltonen:2018tqp,Aaltonen:2007ae,Abazov:2013gpa,Abazov:2008ez}
and the
LHC~\cite{Aad:2014qxa,Aaboud:2017soa,Aad:2013ysa,Aad:2014rta,Aaboud:2017hbk,Khachatryan:2014uva,Khachatryan:2016fue,Sirunyan:2017wgx,Khachatryan:2014zya,Khachatryan:2016crw,Sirunyan:2018cpw,AbellanBeteta:2016ugk},
covering a large range in transverse momentum and rapidity of the final-state particles.
When compared to theory predictions, these measurements provide important tests of
the dynamics of the Standard Model and help to constrain the momentum distributions of partons in the proton.

The study of the forward-rapidity region for this process
is particularly important for our understanding of parton distribution functions (PDFs)
at extremal values of Bjorken-$x$, due to the different kinematic regimes that are probed
compared to the inclusive case.
Owing to the extended rapidity coverage of the LHC experiments,
data is now available for both highly boosted leptons and jets,
giving direct access to these regions in phenomenological studies.

In order to make this connection more concrete, it is instructive to relate
the event kinematics to the Bjorken-$x$ values that are probed.
For a given vector-boson-plus-jet event, one can directly infer the valid range in Bjorken-$x$ values
from the event kinematics at the hadronic centre-of-mass energy $\sqrt{s}$ through
\begin{align}
  \label{eqn:kinematic_constraints_x12}
  x_1 &\ge \frac{1}{\sqrt{s}}\left(m_\rT^V \cdot \re^{+y^V} +\ptjo \cdot \re^{+y^{j1}} \right) , \notag\\
  x_2 &\ge \frac{1}{\sqrt{s}}\left(m_\rT^V \cdot \re^{-y^V} +\ptjo \cdot \re^{-y^{j1}} \right) ,
\end{align}
with $m_\rT^V = \sqrt{{(\ptv)}^2+{m_{V}^2}}$ denoting the transverse mass.
In this equation, $x_1$ and $x_2$ correspond to the momentum fractions
of the incoming partons present in the colliding protons, \ptv and \ptjo are the transverse momenta of the vector boson and the leading-\pt jet,
$m_{V}$ is the invariant mass of the combined system of the decay products of the vector boson and $y_V$ and $y_{j1}$ are the rapidities of the vector boson and the leading jet.
The equality in the above relations holds at Born level.

In general, the smallest $x$ value that can be probed simultaneously ($x_1 \sim x_2$) is
\begin{align}
  x_{\min} = \frac{m_{V+j}^{\min}}{\sqrt{s}} \; ,
\end{align}
which is relevant primarily for data where fiducial cuts are symmetric in rapidity.
Here $m_{V+j}$ is the invariant mass of the vector-boson-plus-jet final state at LO.
In addition, we have the combined kinematic constraint
\begin{align}
  x_1 x_2 \ge \frac{1}{s} \left(m_\rT^{V,\min}+\pt^{j1,\min}\right)^{2} ,
\end{align}
where $m_\rT^{V,\min}$ and $\pt^{j1,\min}$ are the minimum values of the vector boson transverse mass and leading jet \pt admitted by the fiducial cuts.
This constraint is particularly relevant in phase-space regions that are asymmetric in rapidity, which in turn probes more asymmetric values in $x_1$, $x_2$ and gives rise to a more complex interplay between the kinematics and the event selection cuts.

\begin{figure}
  \centering
  \includegraphics[width=0.92\linewidth]{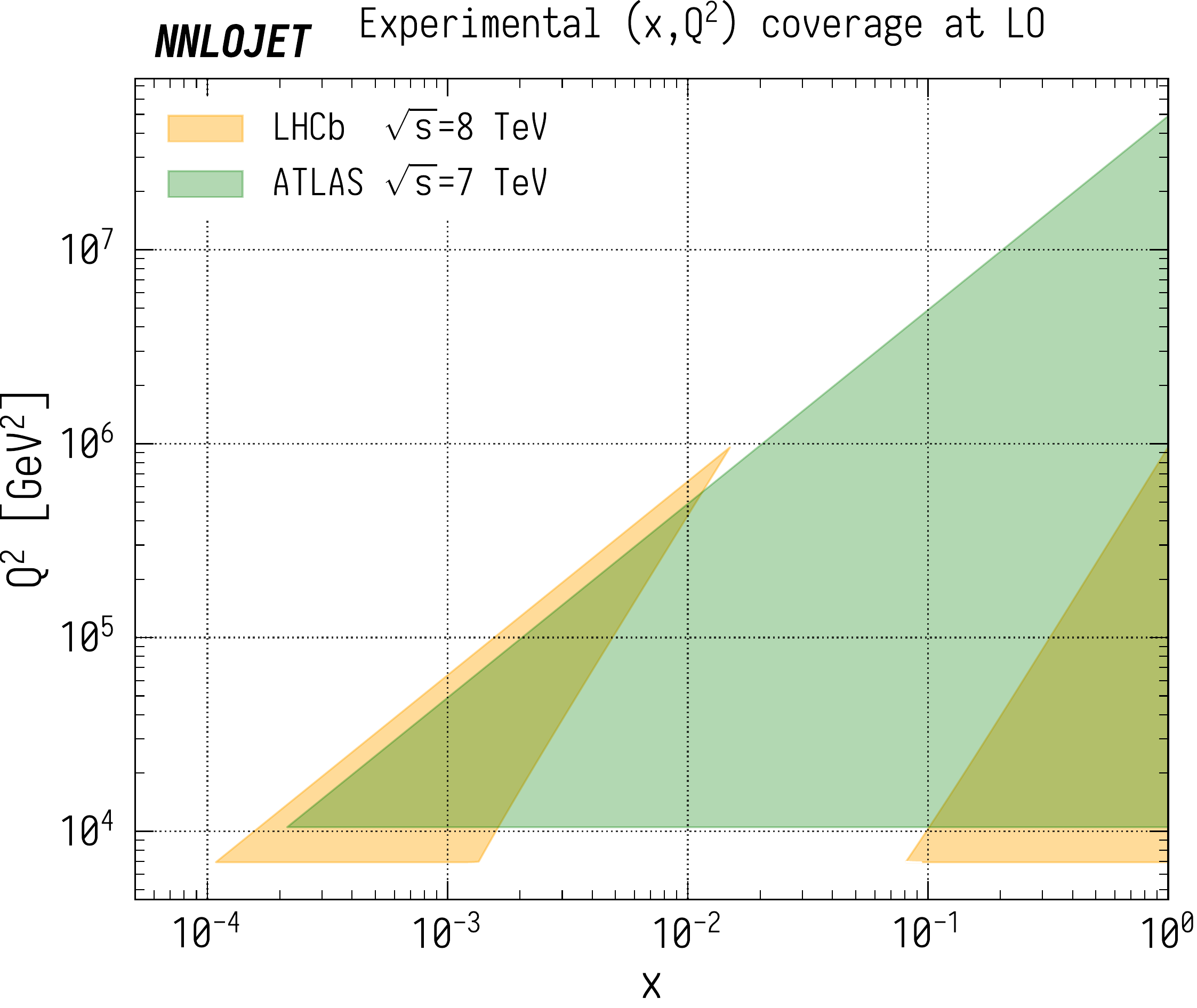}
  \caption{\label{fig:kinematic_xq2}
    The regions of the ($x$, $Q^2$) plane accessible for the LHCb~\cite{AbellanBeteta:2016ugk} and ATLAS~\cite{Aad:2014rta} \PZJ selection criteria at LO. Here $Q^2$ is the invariant mass of the full final state including both charged leptons and QCD radiation and $x$ is the Bjorken-x from either of the incoming beams.}
\end{figure}

Precision QCD predictions for the production of a vector boson in association with a jet have advanced considerably in recent years with the completion of fixed-order next-to-next-to-leading (NNLO) QCD calculations~\cite{Boughezal:2015dva,Ridder:2015dxa,Boughezal:2015ded,Boughezal:2016dtm,Boughezal:2016isb,Ridder:2016nkl,Gehrmann-DeRidder:2016jns,Campbell:2017dqk,Gauld:2017tww,Gehrmann-DeRidder:2017mvr}, which are now being matched to resummation results~\cite{Bizon:2018foh,Sun:2018icb} to extend their validity across a wider kinematic range.
These are complemented by NLO electroweak
corrections~\cite{Denner:2009gj,Denner:2011vu,Kallweit:2015dum}, which are
particularly relevant at large transverse momenta.
There is a strong experimental motivation for precise predictions for these processes due to the high statistics and clean decay channels observed at the LHC, and their relevance to determinations of Standard Model parameters and as
backgrounds for new physics searches~\cite{Lindert:2017olm}.
Fitting procedures for PDFs also benefit greatly from improved predictions,
due to the increased sensitivity to the gluon and quark content of the
proton~\cite{Ball:2017nwa,Boughezal:2017nla}.
Owing to the large gluon luminosity at the LHC, the dominant initial state
for vector-boson-plus-jet production is quark--gluon scattering, with different quark flavour combinations probed
by the different bosons.

\begin{figure}
  \centering
  \includegraphics[width=0.92\linewidth]{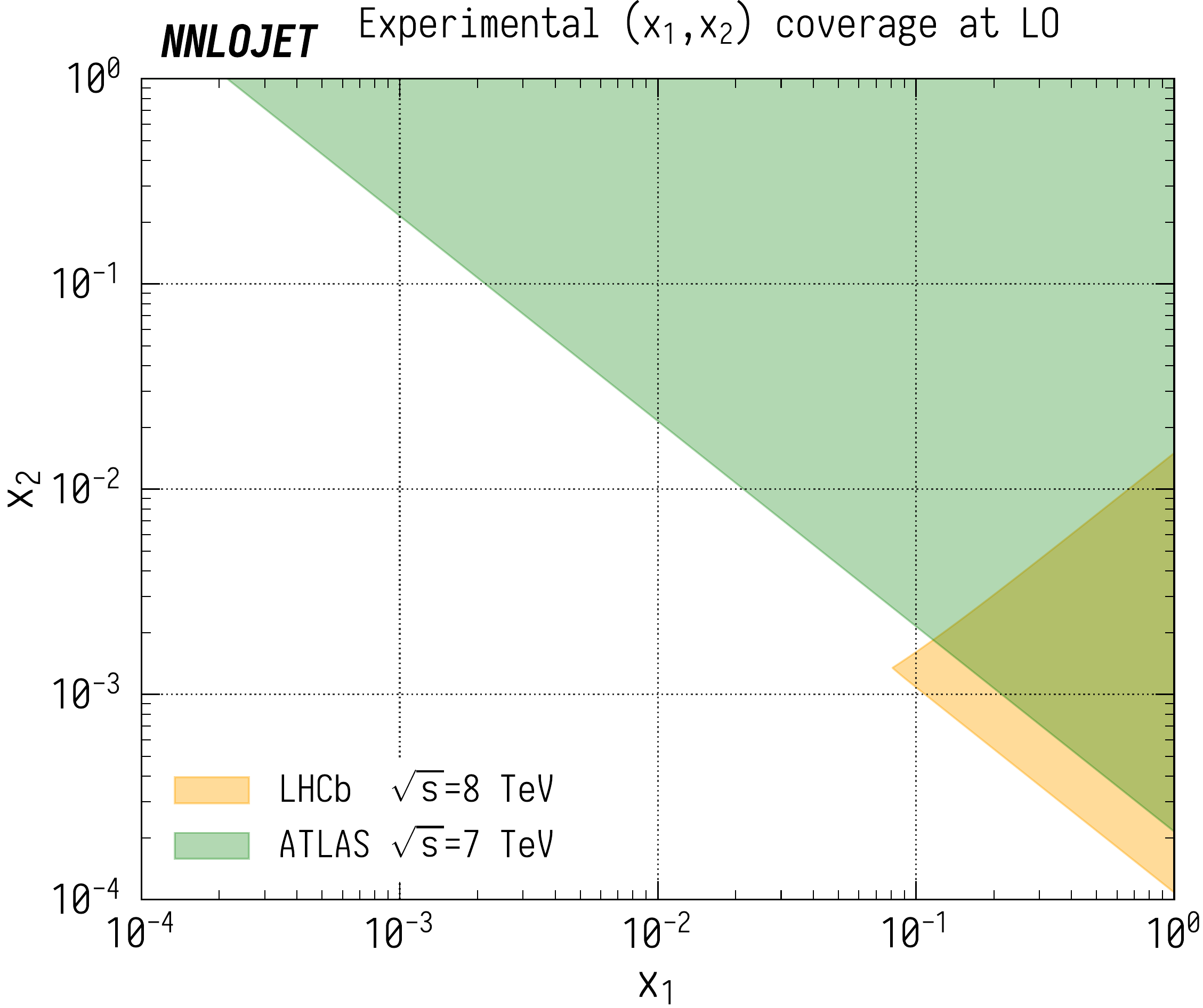}
  \caption{\label{fig:kinematic_x1x2}
    The regions of the ($x_1$, $x_2$) plane accessible for the LHCb~\cite{AbellanBeteta:2016ugk} and ATLAS~\cite{Aad:2014rta} \PZJ selection criteria at LO. Here $x_1$ and $x_2$ are the Bjorken-x values from beams 1 and 2 respectively.}
\end{figure}

In this paper, we perform a comparison between NNLO QCD predictions for vector-boson-plus-jet (VJ) production and measurements by the LHCb~\cite{AbellanBeteta:2016ugk} and ATLAS~\cite{Aad:2014rta} experiments.
These measurements are highly complementary, allowing one to probe a much larger kinematic region than if either of them were taken alone due to the different rapidity coverages of the two detectors. The region of the ($x$, $Q^2$) plane which is probed at LO in QCD in \PZJ production is shown in Fig.~\ref{fig:kinematic_xq2}, where one can see that LHCb covers two distinct sectors corresponding to the $x$ values of the two beams. The corresponding plot for the ($x_1$, $x_2$) plane is shown in Fig.~\ref{fig:kinematic_x1x2}, where the asymmetry of the LHCb region preferentially probes large $x_1$ and small $x_2$ values in contrast to the symmetric ($x_1$, $x_2$) coverage of the ATLAS fiducial region. The kinematic constraints on the LHCb region are relaxed beyond LO as the presence of radiation permits larger $Q^2$ and $x_2$ values, unlike on the ATLAS region where LO kinematics already fully cover the kinematic region accessible at higher orders. The LO kinematics dominates in the contribution to the total cross section however, and gives a good indication of where the sensitivities of the two experiments lie.

The theoretical predictions are obtained using the \NNLOJET framework~\cite{Ridder:2015dxa,Gehrmann-DeRidder:2017mvr}, which implements
the relevant NNLO VJ matrix elements~\cite{Garland:2001tf,Garland:2002ak,Glover:1996eh,Bern:1996ka,Campbell:1997tv,Bern:1997sc,Hagiwara:1988pp,Berends:1988yn} and uses the antenna subtraction method~\cite{GehrmannDeRidder:2005cm,Daleo:2006xa,Currie:2013vh}
to extract and combine infrared singularities from partonic subprocesses with different multiplicity.

Throughout this work, the theoretical predictions employ a diagonal CKM matrix.
The electroweak parameters are set according to the $G_{\mu}$ scheme with the following input parameters:
\begin{align}
  M_\PZ &= 91.1876~\GeV , &
  M_\PW &= 80.385~\GeV , \notag\\
  \Gamma_\PZ &= 2.4952~\GeV , &
  \Gamma_\PW &= 2.085~\GeV , \notag\\
  G_\mathrm{F} &= \mathrlap{ 1.1663787 \times 10^{-5}~\GeV^{-2} ,}
\end{align}
and the PDF set used at all perturbative orders is the central replica of \verb|NNPDF31_nnlo|~\cite{Ball:2017nwa} with $\alphas(M_\PZ)=0.118$.

\section{LHC\MakeLowercase{b} 8~T\MakeLowercase{e}V Boosted Cuts}
\label{sec:LHCb}

At the proton--proton centre-of-mass energy of $8~\TeV$, the LHCb experiment has measured both \PW- and \PZ-boson production in association with a jet with the vector bosons decaying in the muon channel~\cite{AbellanBeteta:2016ugk}.
The acceptance in the forward region of the LHCb experiment allows it to reliably probe PDFs at both much higher and lower momentum fractions $x$ than the general-purpose detectors at the LHC.
This sensitivity arises from kinematic configurations that are asymmetric in $x_1$ and $x_2$, which in turn means that the event is boosted into the forward region.
PDF uncertainties at large $x$ and $Q^2$ are generally driven by uncertainties in the $d$ content of the proton, which these measurements have the capacity to constrain due to their flavour sensitivity, particularly in the charged-current channels.
This provides a strong motivation to use the state-of-the-art NNLO QCD results to test the quantitative agreement of the predictions with the experimental data.

The fiducial cuts applied to the charged leptons and the jets, which we label as the LHCb cuts for both \PWpmJ and \PZJ production are given by
\begin{align}
  \pt^{j} &> 20~\GeV , &
  2.2 &< \eta^{j} < 4.2 , \notag\\
  \pt^{\Pgm} &> 20~\GeV , &
  2 &< y^{\Pgm} < 4.5 , \notag\\
  \Delta R_{\Pgm,j} &> 0.5 ,
\end{align}
where $\pt^{j}$ and $\pt^{\mu}$ are the transverse momenta of the jets and muons respectively, $\eta^{j}$ is the jet pseudorapidity, $y^{\Pgm}$ is the muon rapidity and $\Delta R_{\mathrm{\mu,j}}$ is the angular separation between the leading jet and the muon.
In addition, the requirement $\pt^{\Pgm+j} > 20~\GeV$ is applied for $\PWpmJ$ production, where $\pt^{\Pgm+j}$ is the transverse component of the vector sum of the charged lepton and jet momenta.
For $\PZJ$ production, the invariant mass of the dimuon system $m_{\mu\mu}$ is restricted to the window $60~\GeV < m_{\mu\mu} < 120~\GeV$ around the \PZ-boson resonance.
The anti-$k_\rT$ jet algorithm~\cite{Cacciari:2008gp} is used throughout, with radius parameter $R=0.5$.
In the LHCb analysis~\cite{AbellanBeteta:2016ugk}, the VJ data were compared to NLO theory predictions, which were
observed to overshoot the data throughout, albeit being consistent within the combined theoretical and experimental uncertainties.

For the theoretical predictions presented in this section, we set the central scale as in~\cite{AbellanBeteta:2016ugk}, i.e.,
\begin{equation}
  \mur = \muf =
  \sqrt{{m^2_{V}+\sum\nolimits_i(p^i_{\rT,j})^2}}
  \ \equiv \mu_{0} ,
\end{equation}
with scale variations performed independently for the factorisation and renormalisation scales $\muf$, $\mur$ by factors of $\frac{1}{2}$ and $2$ subject to the constraint $\frac{1}{2}<\muf/\mur<2$.

\begin{table}
  \centering
  \begin{tabular}{c@{\quad}l@{\quad}c}
    \toprule
    \\[-0.8em]
    Process& & Fiducial $\sigma$ [pb] \\
    \hline
    &&\\[-0.8em]
    $\PWpJ$  & LO   & $46.9^{+5.6}_{-2.2}$\\
    &&\\[-0.9em]
      & NLO   & $62.8^{+3.6}_{-3.5}$\\
    &&\\[-0.9em]
      & NNLO   & $63.1^{+0.4}_{-0.5}$ \\
    &&\\[-0.9em]
    & LHCb   & $56.9\pm0.2\pm5.1\pm0.7$\\
    \hline
    &&\\[-0.8em]
    $\PWmJ$  & LO    & $27.2^{+3.2}_{-2.6}$\\
    &&\\[-0.9em]
            & NLO    & $36.7^{+2.2}_{-2.1}$\\
    &&\\[-0.9em]
            & NNLO   & $36.8^{+0.3}_{-0.2}$\\
    &&\\[-0.9em]
    & LHCb   & $33.1\pm0.2\pm3.5\pm0.4$ \\
    \hline
    &&\\[-0.8em]
    $\PZJ$  & LO     & $4.59^{+0.53}_{-0.43}$\\
    &&\\[-0.9em]
            & NLO    & $6.04^{+0.32}_{-3.1}$\\
    &&\\[-0.9em]
            & NNLO   & $6.03^{+0.02}_{-0.04}$\\
    &&\\[-0.9em]
    & LHCb   & $5.71\pm0.06\pm0.27\pm0.07$\\
    \botrule
  \end{tabular}
  \caption{\label{tab:LHCb_fiducial}{Fiducial cross sections for fixed order theory predictions and LHCb results from Ref.~\cite{AbellanBeteta:2016ugk}. The errors quoted for \NNLOJET~ correspond to the scale uncertainty and the reported LHCb errors are statistical, systematic and luminosity respectively.}}
\end{table}

The predictions for the fiducial cross section are shown in Table~\ref{tab:LHCb_fiducial} for LO, NLO and NNLO QCD
and compared to the results reported by the LHCb experiment for the individual VJ channels.
We see large corrections when going from LO to NLO as observed in the NLO/LO K-factor of 1.34 for $\PWm$, $1.35$ for $\PWp$ and 1.32 for
$\PZ$.
On the other hand, going from NLO to NNLO produces much smaller and more stable corrections, with a NNLO/NLO K-factor of $1.006$ for $\PWm$, $1.003$ for $\PWp$ and 0.998 for $\PZ$.
The NNLO corrections lie within the scale bands of the NLO results.
We note that the uncertainty bands overlap marginally between theory and data in Table~\ref{tab:LHCb_fiducial} for $\PWm$ and
$\PZ$ production, but not for $\PWp$ production.

\subsection{Distributions Differential in Leading Jet \pt}

\begin{figure}
  \centering
  \includegraphics[page=13,width=.95\linewidth]{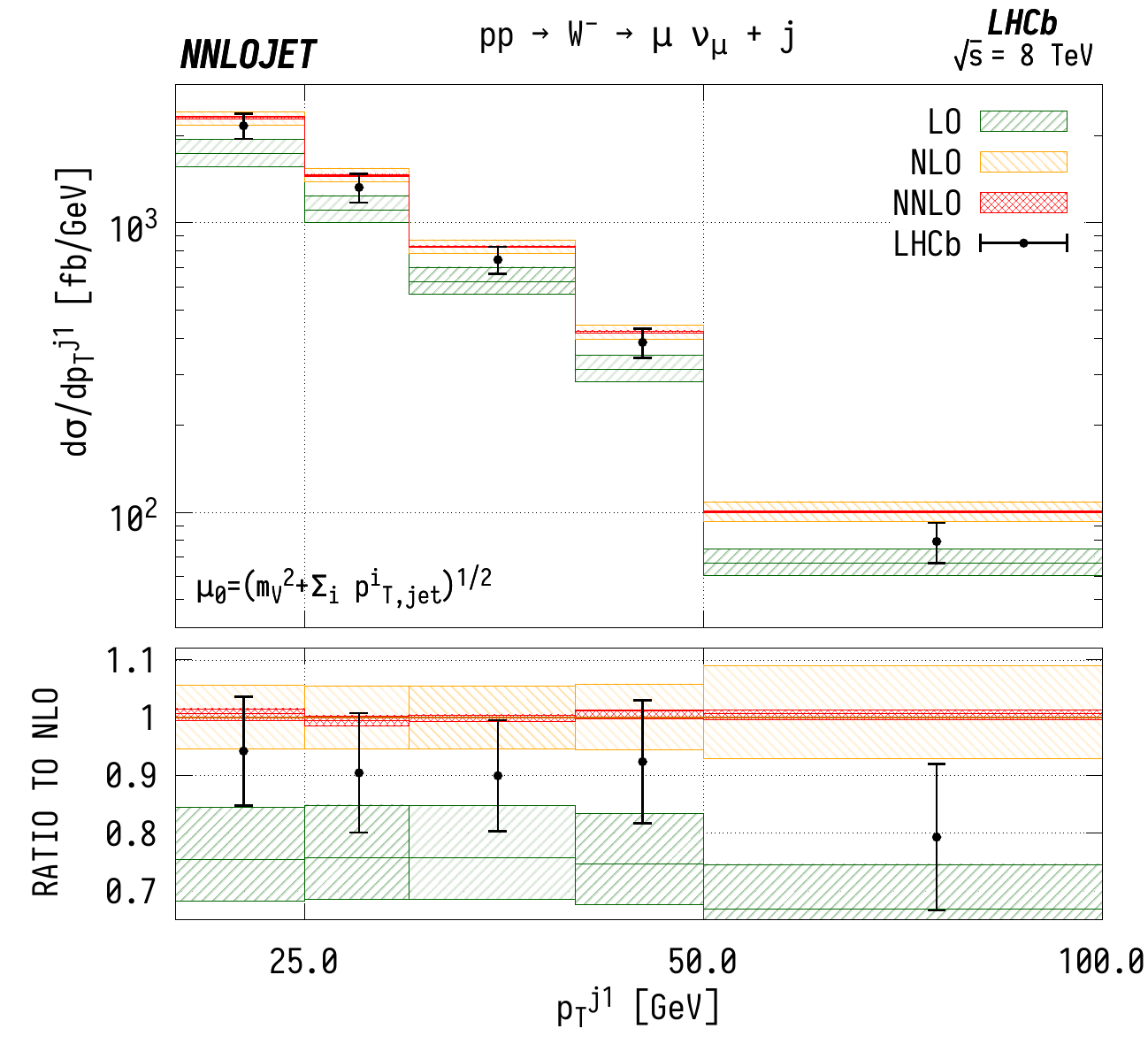}
  \caption{\label{fig:LHCb_Wm_pt}
    Cross section differential in the \pt of the leading jet for \PWm production.
    Predictions at LO (green), NLO (orange), and NNLO (red) are compared to LHCb data from Ref.~\cite{AbellanBeteta:2016ugk}, and the ratio to NLO is shown in the lower panel. The bands correspond to scale uncertainties estimated as described in the main text.
  }
\end{figure}
\begin{figure}
  \centering
  \includegraphics[page=14,width=.95\linewidth]{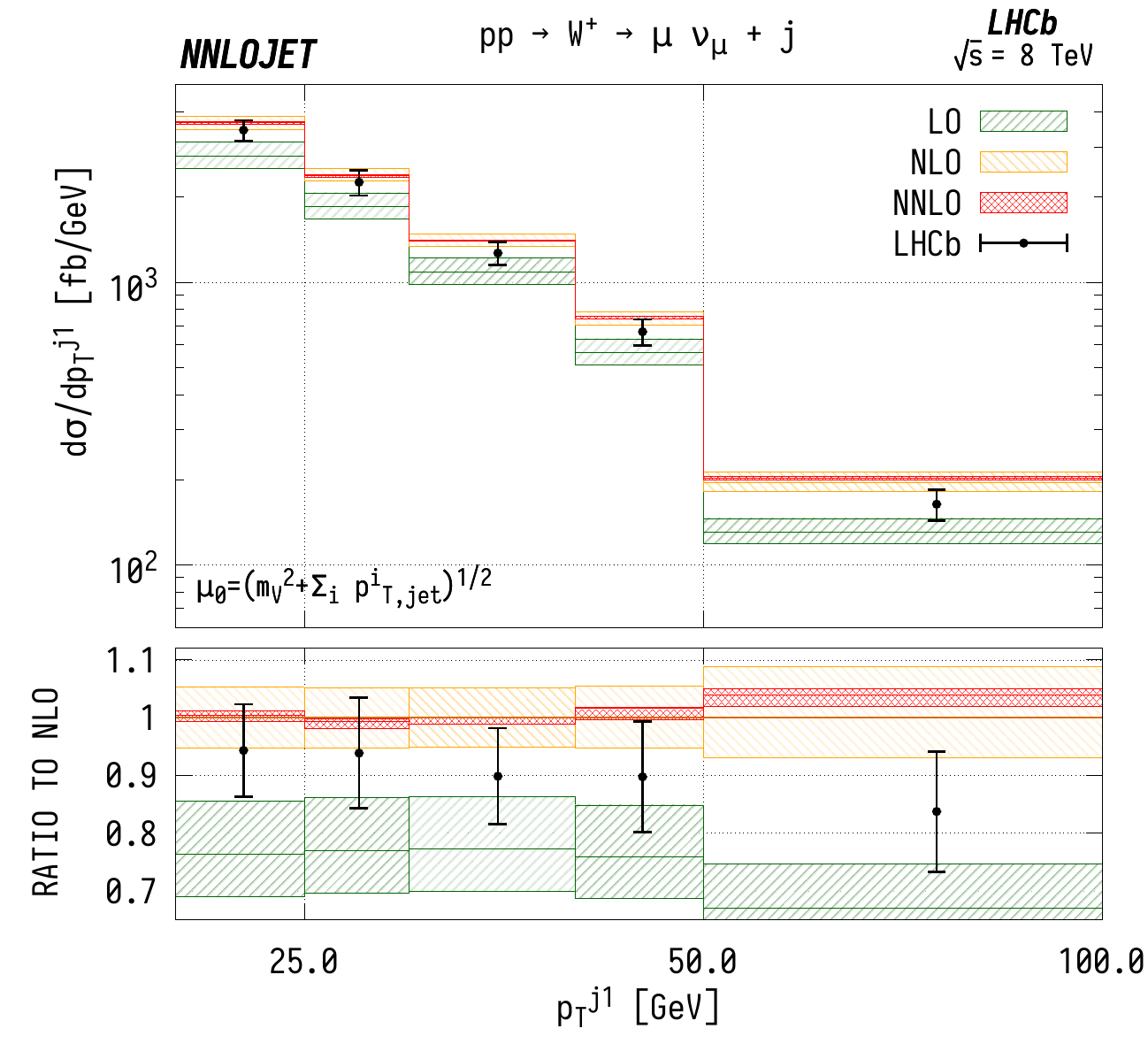}
  \caption{\label{fig:LHCb_Wp_pt}
    Cross section differential in the \pt of the leading jet for \PWp production. See Fig.~\ref{fig:LHCb_Wm_pt} for details.}
\end{figure}
\begin{figure}
  \centering
  \includegraphics[page=15,width=.95\linewidth]{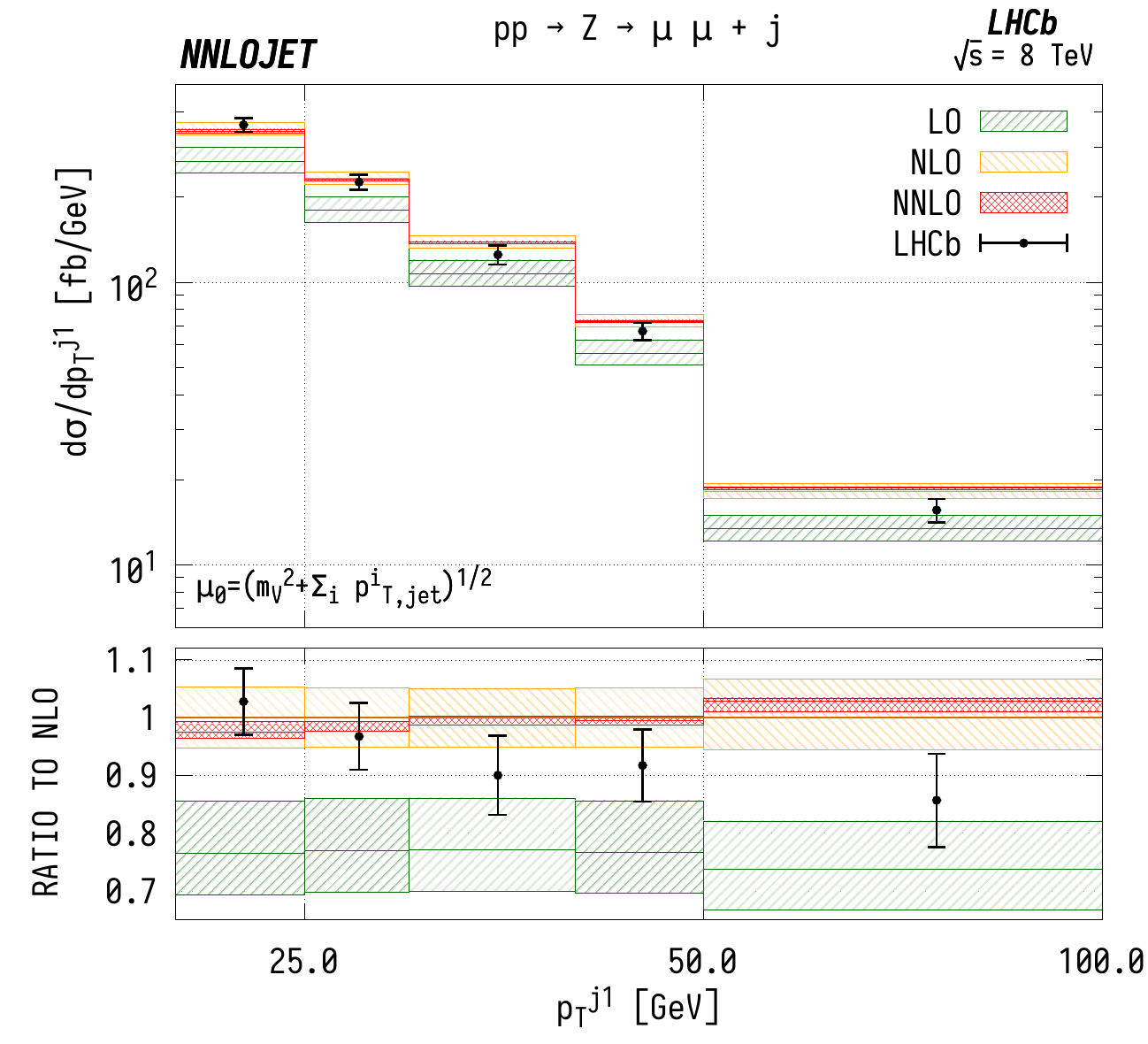}
  \caption{\label{fig:LHCb_Z_pt}
    Cross section differential in the \pt of the leading jet for \PZ production.  See Fig.~\ref{fig:LHCb_Wm_pt} for details.
  }
\end{figure}

Figures~\ref{fig:LHCb_Wm_pt}--\ref{fig:LHCb_Z_pt} show the distributions for transverse momentum of the leading jet in $\PWm$, $\PWp$ and $\PZ$ production respectively.
Similarly to the fiducial cross section, the scale dependence of the differential distributions is considerably reduced when going from NLO to NNLO.
The NNLO corrections are stable with respect to NLO, indicating a good convergence of the perturbative series.
In addition, these results exhibit a strong similarity in behaviour between the $\PWm$, $\PWp$ and $\PZ$ production channels.
We see that the theory overshoots the data by $\sim 5$--$10\%$ over the bulk of the distribution, rising to $30\%$ in the highest \pt bin.
This closely mirrors the effects seen at NLO as well as in the total cross section.
The considerable decrease in theory uncertainty from NLO to NNLO makes the tension between data and theory more pronounced.

 \begin{figure}
  \centering
  \includegraphics[page=7,width=.95\linewidth]{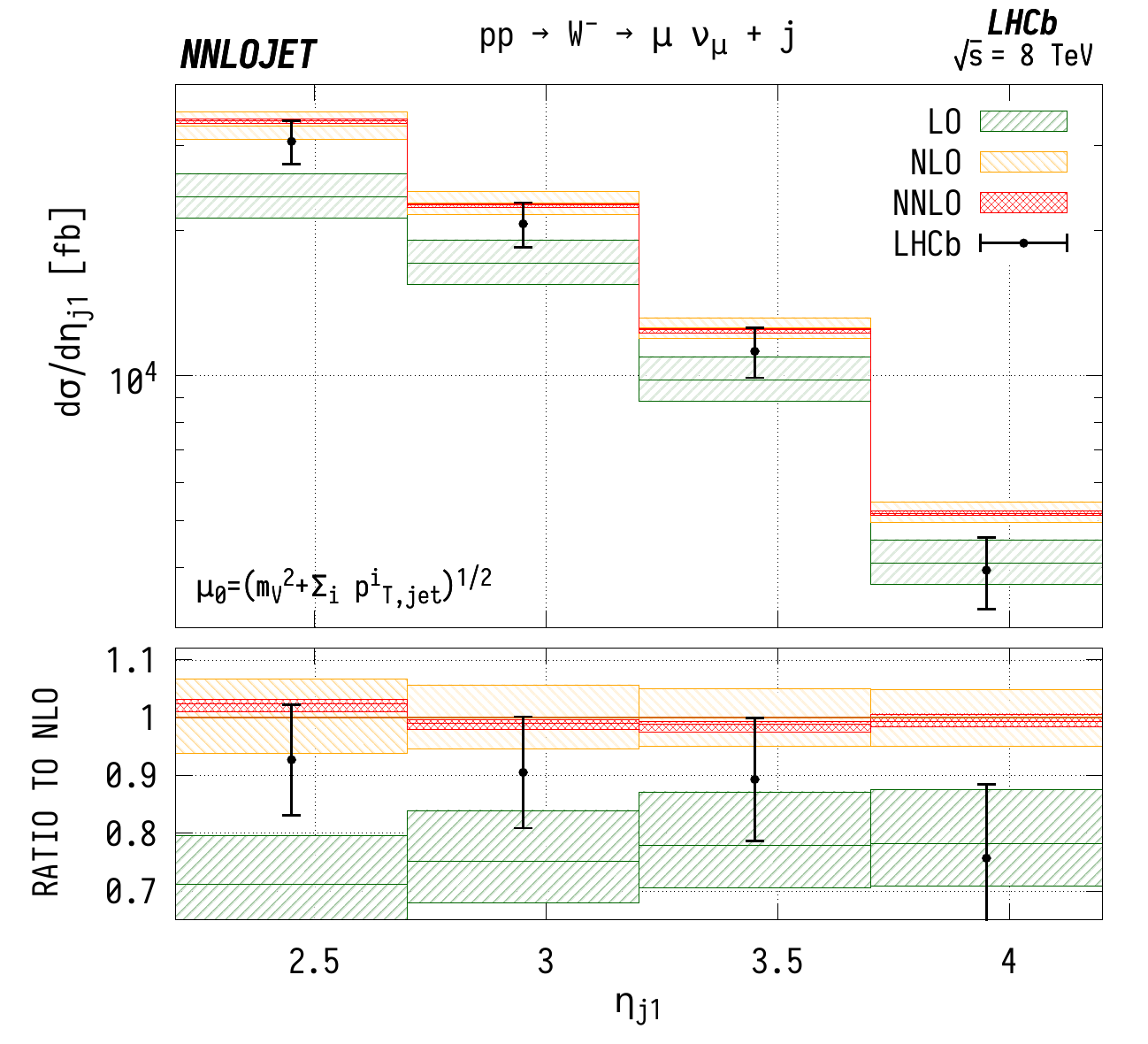}
  \caption{\label{fig:LHCb_Wm_etaj1}
    Cross section differential in the pseudorapidity $\eta$ of the leading jet for $\PWm$ production.
     See Fig.~\ref{fig:LHCb_Wm_pt} for details.
  }
\end{figure}
\begin{figure}
  \centering
  \includegraphics[page=8,width=.95\linewidth]{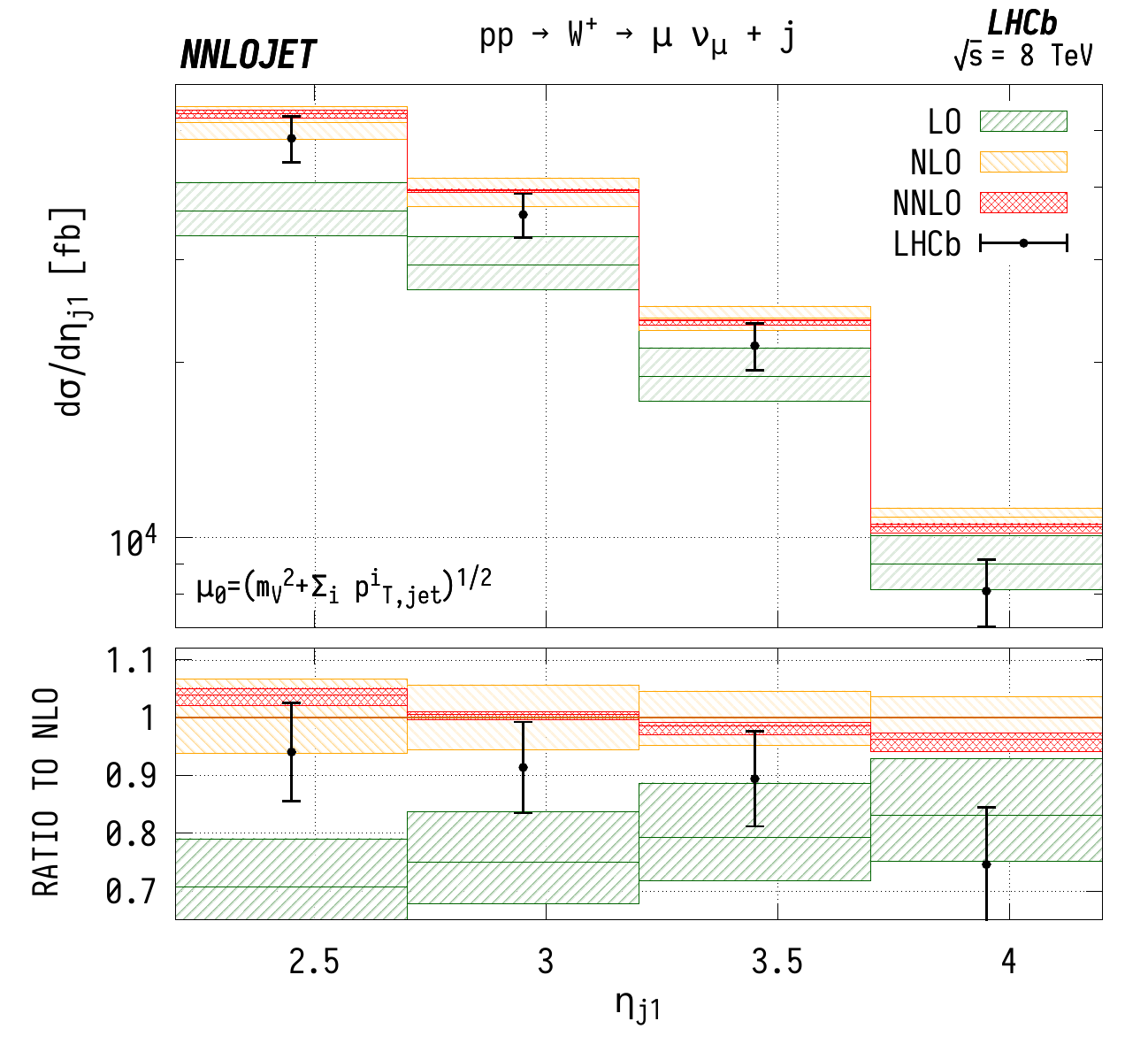}
  \caption{\label{fig:LHCb_Wp_etaj1}
    Cross section differential in the pseudorapidity $\eta$ of the leading jet for $\PWp$ production.
     See Fig.~\ref{fig:LHCb_Wm_pt} for details.
  }
\end{figure}
\begin{figure}
  \centering
  \includegraphics[page=9,width=.95\linewidth]{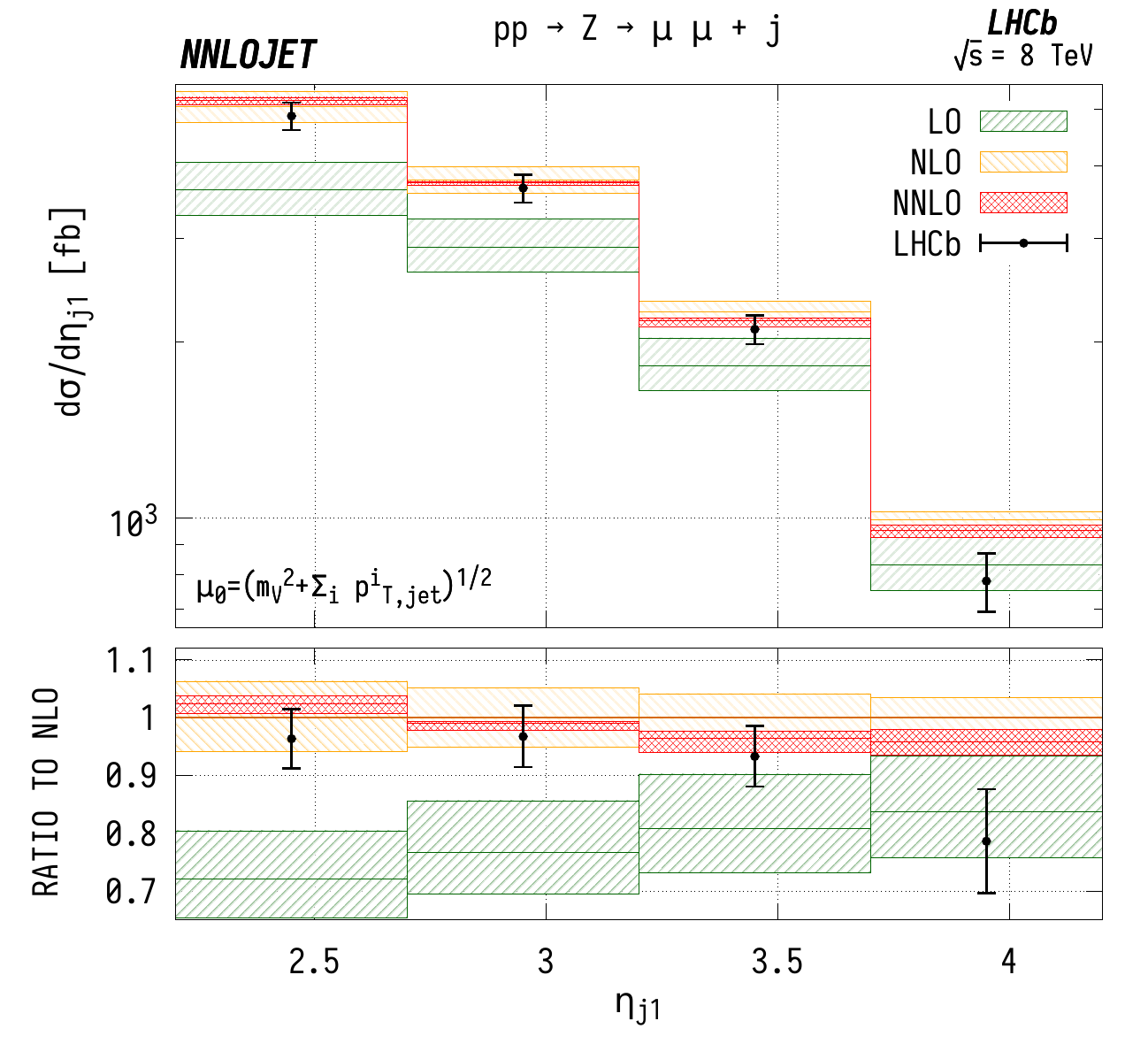}
  \caption{\label{fig:LHCb_Z_etaj1}
    Cross section differential in the pseudorapidity $\eta$ of the leading jet for $\PZ$ production.
     See Fig.~\ref{fig:LHCb_Wm_pt} for details.
  }
\end{figure}

For the cuts placed on the \PWJ final state, we are also able to associate the bins in \ptj to lower limits on the Bjorken-$x$ invariants.
The lowest \pt bin has the loosest constraint on the forward $x$, with $x_1>0.041$, $x_2>5.4\times 10^{-5}$.
However, for the highest \pt bins, between $50$ and $100~\GeV$, the restrictions translate to $x_1>0.075$, $x_2>0.00011$.
Due to the invariant mass cuts applied in the \PZJ case shown in Fig.~\ref{fig:LHCb_Z_pt}, the smallest values in Bjorken-$x$ that can be probed only extend down to $x_1 > 0.11$, $x_2 > 0.0002$ in the highest \pt bin.
As a result, one probes larger values of $x$ for \PZJ production than for \PWJ in general. At large \pt, we see that the same features are present in the neutral and charged current cases.
We observe that the NNLO predictions overshoot the data.

\subsection{Distributions Differential in Pseudorapidity}

The leading jet pseudorapidity distributions in Figs.~\ref{fig:LHCb_Wm_etaj1}--\ref{fig:LHCb_Z_etaj1} show a similar pattern of deviation between NNLO predictions and data to the previous \ptj results,
with theory predictions exceeding the data at the largest values of $\eta_{j1}$.
The behaviour is similar for  $\PWp$, $\PWm$ and $\PZ$, which may further indicate that the discrepancy is mainly due to the gluon distribution being overestimated at large $x$.
Changes in individual quark or antiquark distributions would instead give a pattern of
discrepancy that is more pronounced in one of the channels than in the others.
In the pseudorapidity distributions, we probe simultaneously more extreme regions of $x_1$ and $x_2$ than for the \ptj distributions as the directional dependence on $y_{j}$ as given in Eq.~\eqref{eqn:kinematic_constraints_x12} allows us to more directly discriminate the two Bjorken-$x$ values.
This can be seen most explicitly for the \PZJ case, for which the forward-most bin in pseudorapidity requires implicitly $x_1>0.16$, $x_2>1.1\times 10^{-4}$, meaning that the large $x>\cO(0.1)$ regions are probed efficiently.

\begin{figure}
  \centering
  \includegraphics[page=10,width=.95\linewidth]{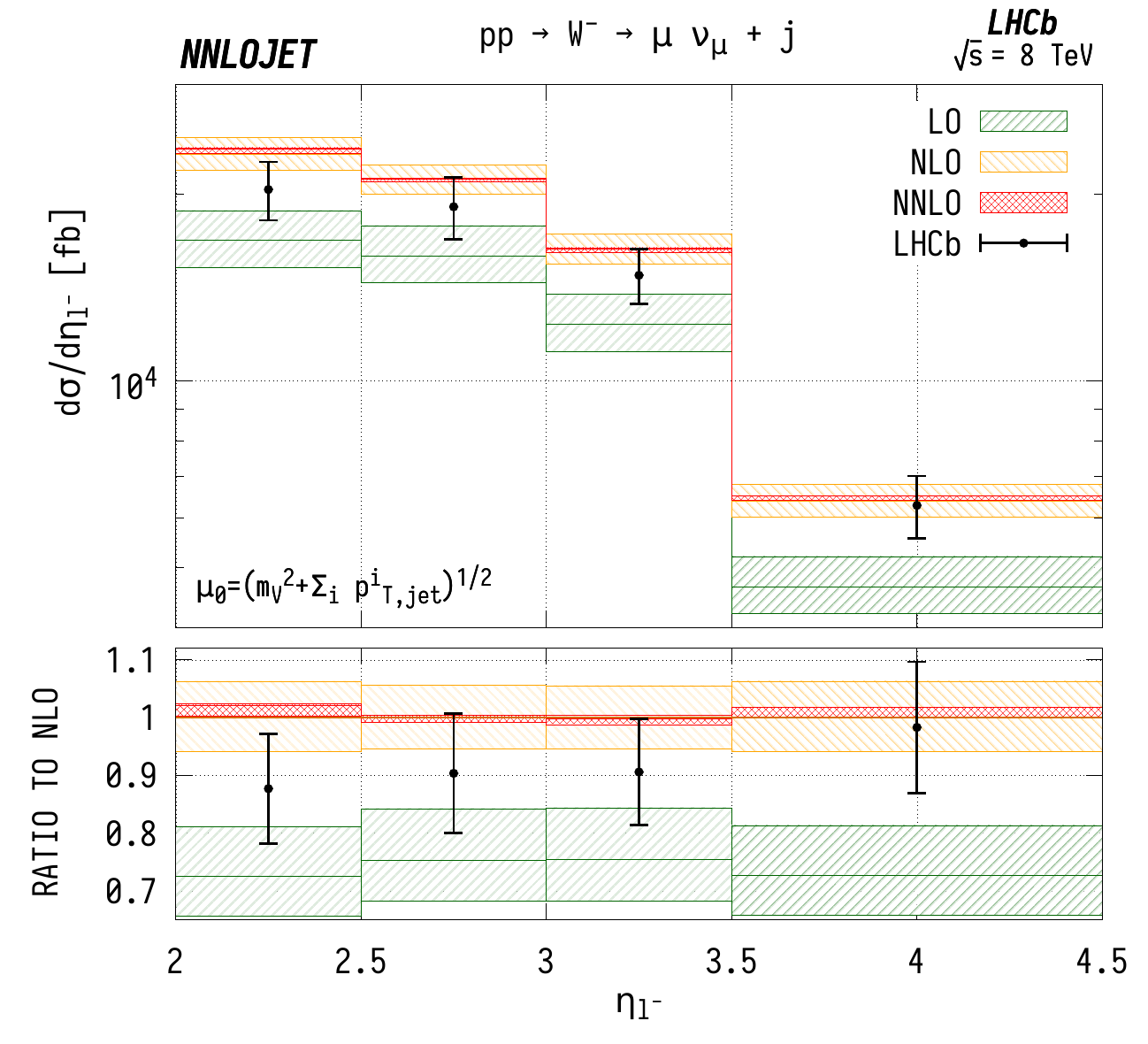}
  \caption{\label{fig:LHCb_Wm_etalep}
    Cross section differential in the pseudorapidity $\eta$ of the lepton for $\PWm$J production.
    See Fig.~\ref{fig:LHCb_Wm_pt} for details.
  }
\end{figure}
\begin{figure}
  \centering
  \includegraphics[page=11,width=.95\linewidth]{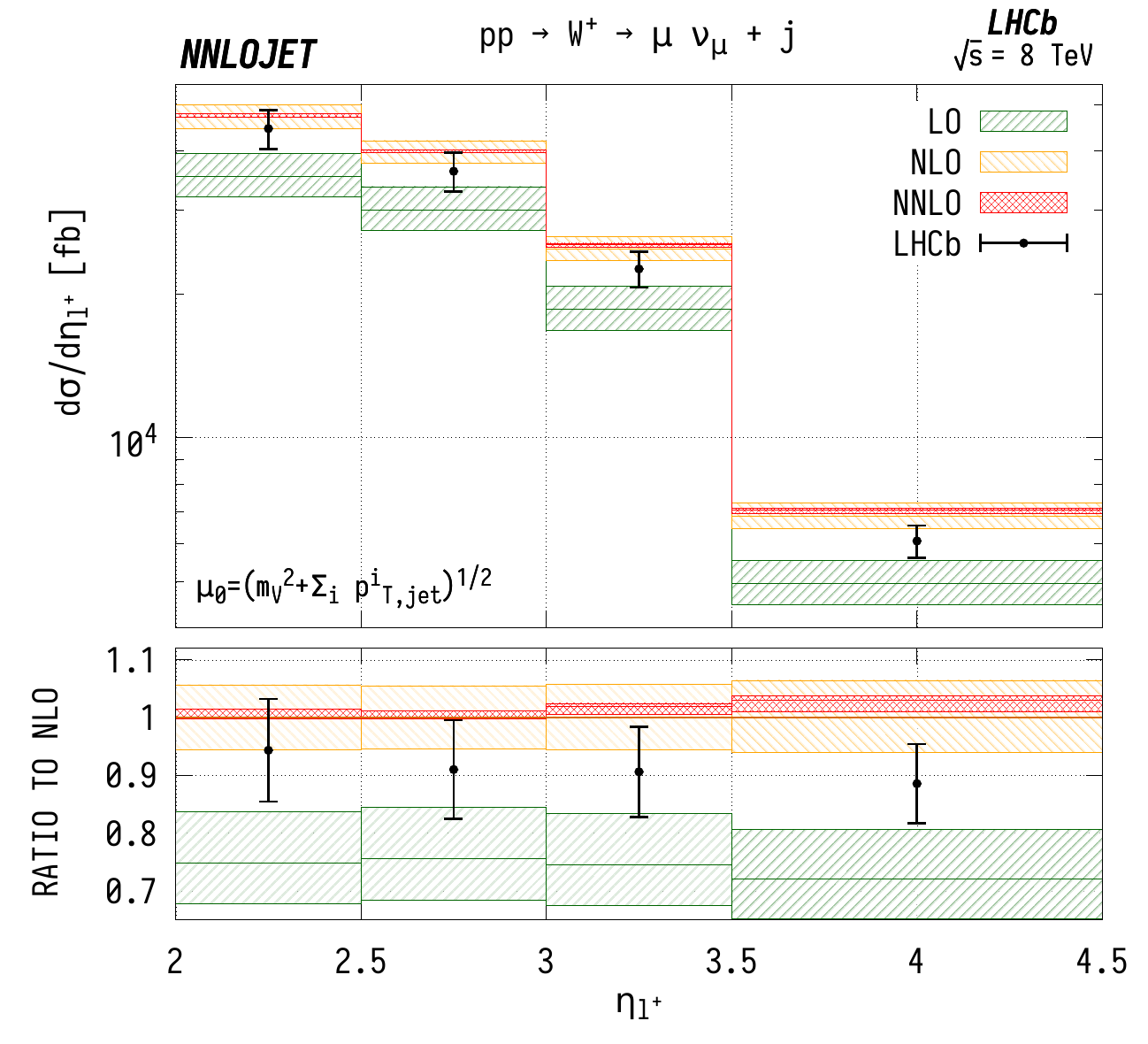}
  \caption{\label{fig:LHCb_Wp_etalep}
    Cross section differential in the pseudorapidity $\eta$ of the lepton for $\PWp$J production. See Fig.~\ref{fig:LHCb_Wm_pt} for details.
  }
\end{figure}
\begin{figure}
  \centering
  \includegraphics[page=12,width=.95\linewidth]{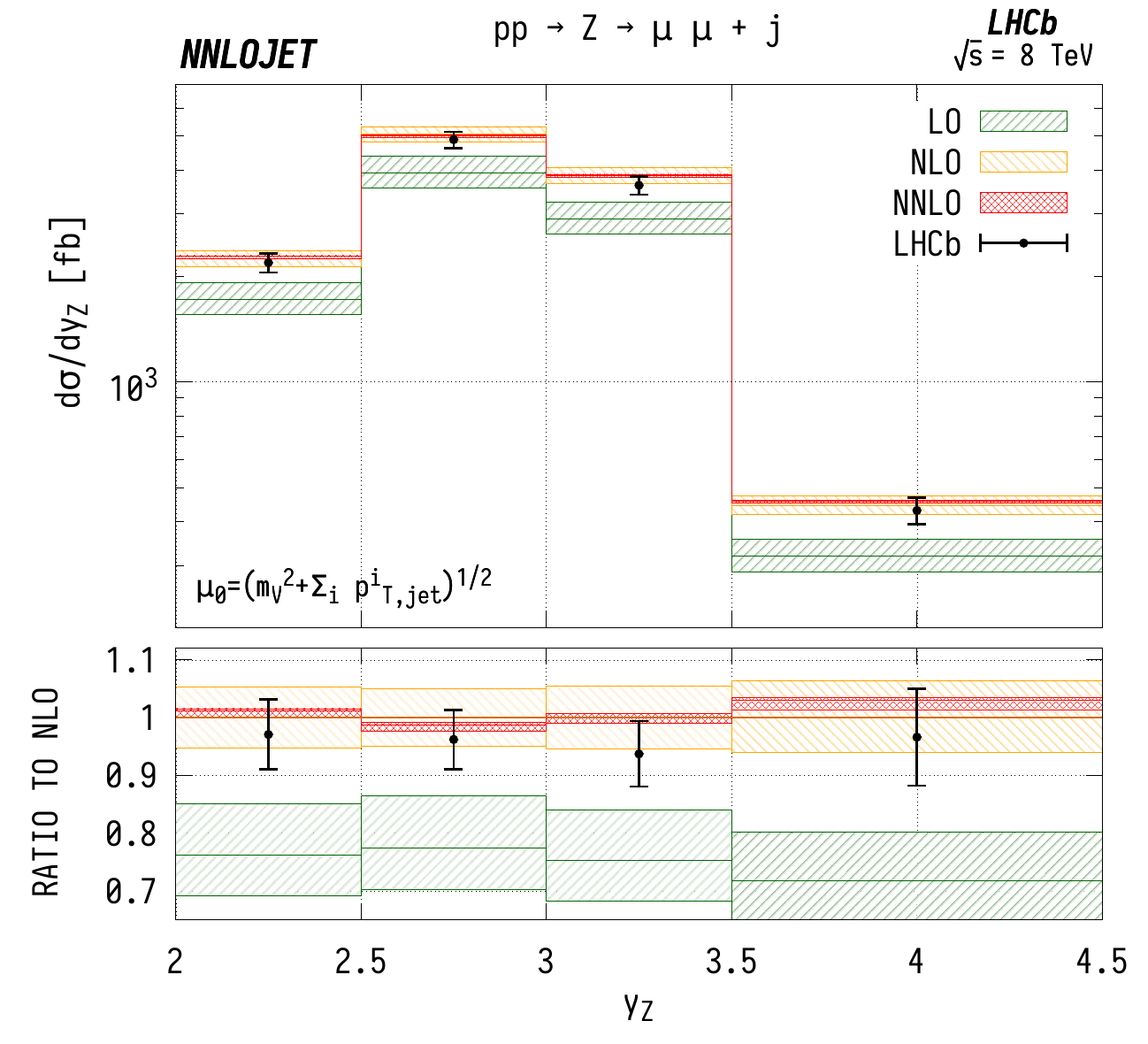}
  \caption{\label{fig:LHCb_Z_yz}
    Cross section differential in the rapidity of the dilepton system for $\PZ$J production. See Fig.~\ref{fig:LHCb_Wm_pt} for details.
  }
\end{figure}

The distributions for the rapidity of the charged lepton $\eta_\ell$ are shown in Figs.~\ref{fig:LHCb_Wm_etalep} and \ref{fig:LHCb_Wp_etalep} for $\PWm$ and $\PWp$ respectively.
Here the NNLO predictions lie $\sim 5$--$15\%$ above the data across the entire considered range in $\eta_\ell$.
Note that it would be preferable to construct these distributions as a function of the \PW rapidity $y_\PW$, which however can not be unambiguously reconstructed experimentally due to the unknown longitudinal component of the neutrino momentum.
For the case of neutral-current production, on the other hand, this is possible and is shown in Fig.~\ref{fig:LHCb_Z_yz} differentially with respect to the rapidity of the reconstructed \PZ boson.

From the charged-current data one can further construct the charge asymmetry differentially in the lepton pseudorapidity $A^{\pm}(\eta_\ell)$,
 \begin{equation}
    A^{\pm}(\eta_\ell) =
    \frac{\rd\sigma^{\PWp j}/\rd\eta_\ell - \rd\sigma^{\PWm j}/\rd\eta_\ell}
         {\rd\sigma^{\PWp j}/\rd\eta_\ell + \rd\sigma^{\PWm j}/\rd\eta_\ell} \; .
    \label{eqn:charge_asym_def}
 \end{equation}
The charge asymmetry is a valuable input to PDF fits as many systematic experimental errors cancel due to correlations in luminosity and systematic errors between the measurements of \PWpJ and \PWmJ, giving a higher level of precision than for the total cross sections alone.
This is also true for the theory predictions, where many higher-order contributions cancel between \PWpJ and \PWmJ, and the similarity of the two calculations justifies some correlation between scale errors.
$A^{\pm}$ directly provides information on the difference between the $u$ and $d$ quark (as well as between the $\bar d$ and $\bar u$ anti-quark) content of the proton.

The advantage of considering the charge asymmetry for events where a jet is produced in association with the \PW boson, which can be regarded as an exclusive asymmetry, as opposed to the inclusive $A^{\pm}$ is that the implicit constraint on Bjorken-$x$ is tightened due to the increase in partonic energy required.
Before comparing our predictions with LHCb data for the exclusive charge asymmetry,  it is instructive to recall the status of measurements of its inclusive analogue.
The LHCb measurement of the inclusive charge asymmetry~\cite{Aaij:2015zlq} probes larger values of $x$ than at ATLAS or CMS.
Currently the main constraints on $u$ and $d$ content at $x>0.1$ come primarily from fixed-target DIS experiments and the  D0 inclusive lepton charge asymmetry data~\cite{D0:2014kma}.
The inclusion of the latest Tevatron results in PDF fits generally results in a harder $u/d$ behaviour in this high-$x$ region~\cite{Dulat:2015mca}.

\begin{figure}
  \centering
  \includegraphics[page=17,width=.95\linewidth]{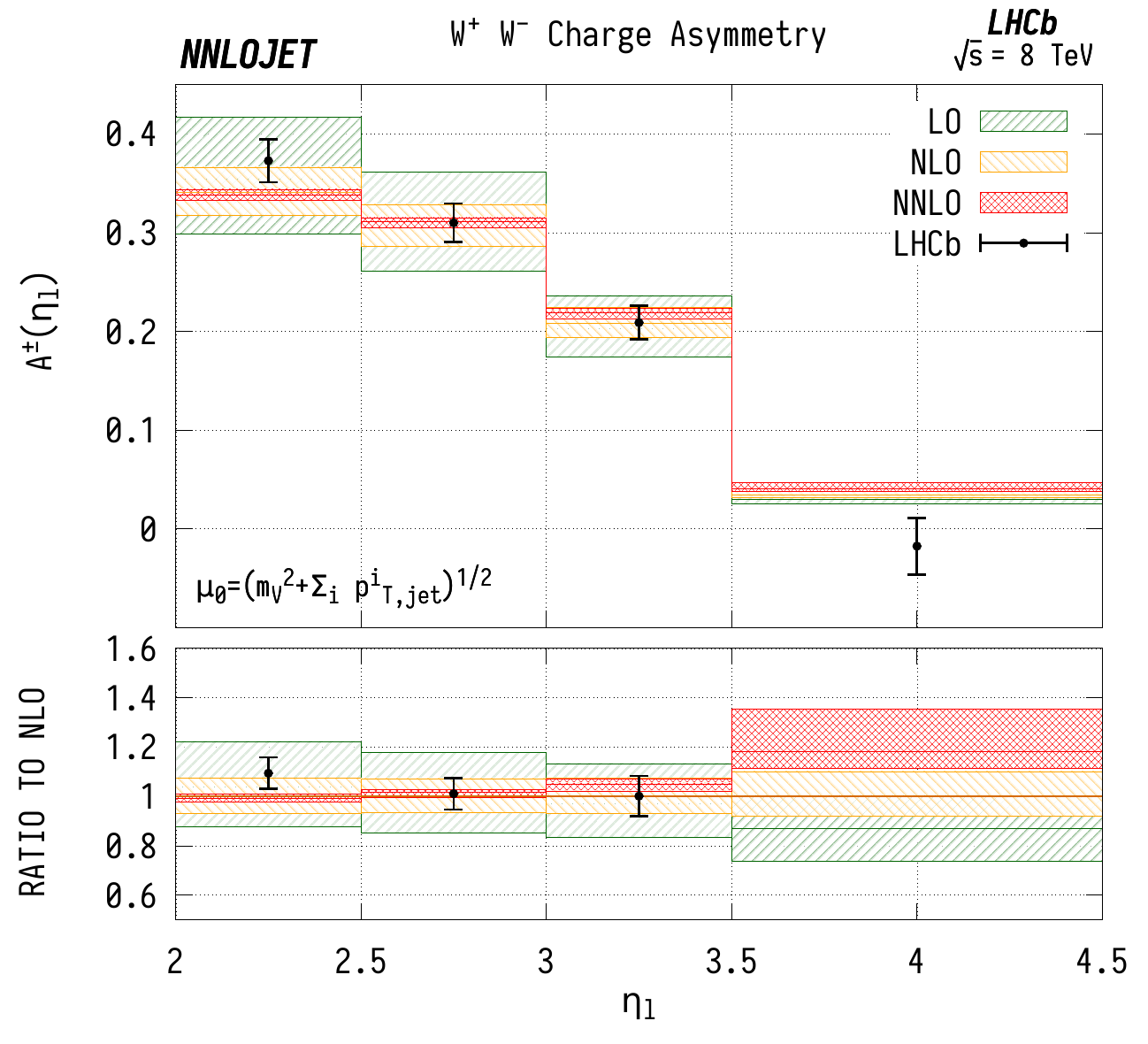}
  \caption{\label{fig:LHCb_Asym_etalep}
    $\PWpm$ asymmetry in WJ production differential in the pseudorapidity $\eta$ of the lepton produced from the \PW boson decay. See Fig.~\ref{fig:LHCb_Wm_pt} for details.
  }
\end{figure}

In Fig.~\ref{fig:LHCb_Asym_etalep}, we show a comparison between our theoretical predictions for $A^{\pm}$ related to WJ production and the LHCb data.
Inside the numerator and the denominator expressions,
we fully correlate the scales between the $\PWp$ and $\PWm$ cross sections, which amounts to taking the sum and difference of the cross sections as independent
physical quantities $\left[\rd\sigma^{\PWp}\pm\rd\sigma^{\PWm}\right](\muf, \mur)$ instead of the $\PWp$ and $\PWm$ cross sections.
The scale uncertainty shown is then obtained by independently varying the factorisation $(\muf)$ and renormalisation $(\mur)$ scales of both the numerator and denominator by factors of $\frac{1}{2}$ and $2$ around the central scale, while imposing the restriction $\frac{1}{2}\leq \mu/\mu'\leq 2$ between all pairs of scales ($\mu,\mu'$) in Eq.~\eqref{eqn:charge_asym_def}.

The shape of $A^{\pm}$ as a function of $\eta_\ell$ is generally determined by two competing effects~\cite{Farry:2015xha}.
The first is the (anti-)quark content of the PDF, where the $u/d$ ratio and $q/\bar{q}$ asymmetry increase with momentum fraction $x$, and therefore with $\eta_\ell$.
This alone gives an increase in $A^{\pm}$ with $\eta_\ell$ since $u$-initiated production is dominant in $\PWp$ production while $d$-initiated production is dominant for $\PWm$. 

\begin{figure}
  \centering
  \includegraphics[page=1,width=.95\linewidth]{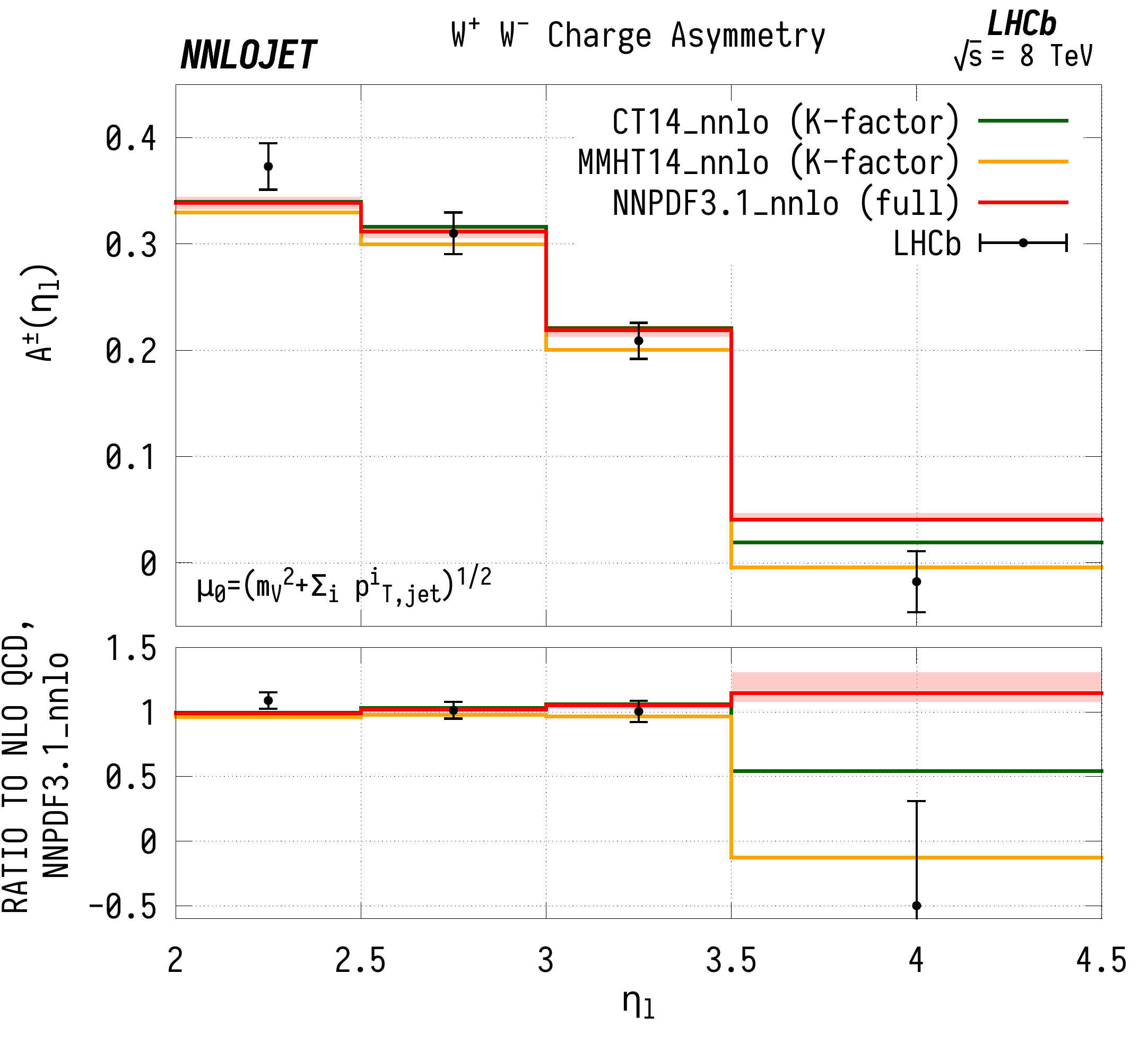}
  \caption{\label{fig:LHCb_pdf_variation}
    $\PWpm$ asymmetry in WJ final states differential in the pseudorapidity $\eta$ of the lepton produced from the \PW~boson decay, evaluated with NNPDF3.1 (red), MMHT14 (yellow), CT14 (green) NNLO
    parton distribution functions. The NNPDF3.1 curve corresponds to a full NNLO calculation with scale uncertainties as described in the main text, and is used to determine a differential NNLO/NLO K-factor.
    The other two predictions are
    calculated at NLO and then rescaled by this K-factor.}
    \end{figure}
The second factor is due to the left-handedness of the couplings in the $\PWpm$ production and decay process, which
results in opposite preferential directions of the positive and negative decay leptons relative to the  $\PWpm$ spin.
As a consequence,
for the $\PWp$ case, the lepton is preferentially produced at lower $\eta$ than the $\PWp$, whereas for the $\PWm$ case, the lepton is preferentially produced at higher relative $\eta$.
This effect causes the asymmetry to decrease with $\eta_\ell$, and dominates over the quark PDF effects
at higher $x$, as can be seen in Fig.~\ref{fig:LHCb_Asym_etalep}.

We find that the NNLO predictions for the asymmetry describe the data reasonably well, but in general show a less steep slope with $\eta_\ell$ than the data.
This may be indicative of a PDF overestimate in the $u/d$ ratio for $x\gtrsim 0.1$ which would lead to the observed overprediction of the charge asymmetry in this region. It is noted that the large $u/d$ ratio is in particular
inferred~\cite{Ball:2017nwa,Dulat:2015mca} from
the Tevatron D0 lepton charge asymmetry data~\cite{D0:2014kma}. It will thus be crucial to combine
these data with the LHCb results~\cite{AbellanBeteta:2016ugk} in a global fit to determine whether they are mutually consistent.

The sensitivity of the $\PWpm$ asymmetry in WJ final states on the PDF parametrizations is illustrated in Figure~\ref{fig:LHCb_pdf_variation}, which shows this asymmetry at NNLO  for
NNPDF3.1~\cite{Ball:2017nwa}, MMHT14~\cite{Harland-Lang:2014zoa} and
CT14~\cite{Dulat:2015mca} parton distributions.
The NNPDF3.1 prediction is obtained from a full NNLO calculation of the individual cross sections entering into the ratio, which are also used to extract NNLO K-factors.
Predictions for the other two PDF parametrizations are computed at NLO at cross section level, and then rescaled
by these K-factors, before computing the ratio. The large spread of the predictions (noting also the different scale
in the ratio compared to Figure~\ref{fig:LHCb_Asym_etalep}) in the last bin reflects the different modelling of the quark distributions at large $x$ in the three parametrizations, and
demonstrates the discriminating power of the LHCb asymmetry measurement.

\section{ATLAS 7~T\MakeLowercase{e}V Standard Cuts}
\label{sec:ATLAS}

The second set of experimental data we consider is the $7~\TeV$ (electron and muon) measurement by the ATLAS experiment~\cite{Aad:2014rta}, which combines data from the \PW and \PZ analyses of~\cite{Aad:2014qxa} and \cite{Aad:2013ysa} with a small modification to the lepton selection criteria applied in the \PZ analysis when taking ratios. This modification is applied in order to better match the \PW selection criteria.

The ATLAS detector has a large rapidity range, capable of measuring pseudorapidities of up to $|\eta|=4.9$ in the endcap region for both hadronic and electromagnetic final states.
Unlike the LHCb measurement region, the large pseudorapidity reach of ATLAS also allows to probe large rapidity separations between final state particles, which correspond to configurations in which the Bjorken-$x$ of both incoming protons is relatively large.
In the following, we perform a comparison of fixed-order NNLO results to the individual \PWJ and \PZJ distributions of~\cite{Aad:2014qxa} and \cite{Aad:2013ysa}, before constructing the ratios of $\PWJ$ ($\equiv\PWpJ+ \PWmJ$) and \PZJ distributions and comparing those to the results of~\cite{Aad:2014rta}.
We consider leading jet \pt distributions in inclusive (at least one jet is required) and exclusive (exactly one jet is required) jet production, as well as inclusive leading jet rapidity distributions.
The inclusive distributions have previously been compared to NNLO QCD predictions
in~\cite{Boughezal:2016dtm}, however exclusive distributions and ratios of distributions were not considered.

The fiducial cuts used in the ATLAS analyses are as follows:
\begin{align}
  \pt^{j} &> 30~\GeV , &
  |y^{j}| &< 4.4 , \notag\\
  \pt^{\ell} &> 25~\GeV , &
  |y^{\ell}| &< 2.5 , \notag\\
  \Delta R_{\ell,j} &> 0.5 \label{eqn:atlas_cuts}.
\end{align}
For $\PWpmJ$ production, the restrictions $E_{\rT}^\miss>25~\GeV$, and $m_{\rT}^\PW>40~\GeV$ on the missing transverse energy and transverse mass of the \PW boson are imposed.
For \PZJ production the requirements $66~\GeV < m_\rT^{\ell\ell} < 116~\GeV$ and $\Delta R_{\ell\ell} > 0.2$ are applied to the transverse mass of the dilepton system and angular separation of the leptons.
In the \PZJ distributions, we relax the lepton \pt cut from $25$ to $20~\GeV$ in order to compare directly with the results of~\cite{Aad:2013ysa}.
However we keep the lepton \pt cut at $25~\GeV$ when constructing ratios of WJ and ZJ distributions.

Jets are reconstructed using the anti-$k_\rT$ algorithm~\cite{Cacciari:2008gp}
with radius parameter $R=0.4$, and we choose the central scale of the theory predictions as
\begin{align}
  \muf = \mur = \frac{1}{2}H_{\rT} = \frac{1}{2}\sum_{i \,\in\, \mathrm{jets,\,\ell,\,\nu}}\pt^i
  \equiv\mu_{0} ,
\end{align}
where $H_\rT$ is the scalar sum of the transverse momenta of all final state jets and leptons/neutrinos as appropriate.
We denote the number of jets as $N$, such that in the selection criteria $N=1$ corresponds to the exclusive case and $N\geq1$ corresponds to the inclusive case.

The scale variation uncertainties for the ratios are obtained in a similar manner as for LHCb \PWpm asymmetries, with fully correlated scales between the \PWp and \PWm processes in the numerator, but taking the envelope of the scales when taking the ratio to the \PZ distributions, imposing $\frac{1}{2}\leq \mu/\mu'\leq 2$ between all pairs of scales.

\subsection{Exclusive \ptjo Distributions}

\begin{figure}
  \centering
  \includegraphics[page=3,width=.95\linewidth]{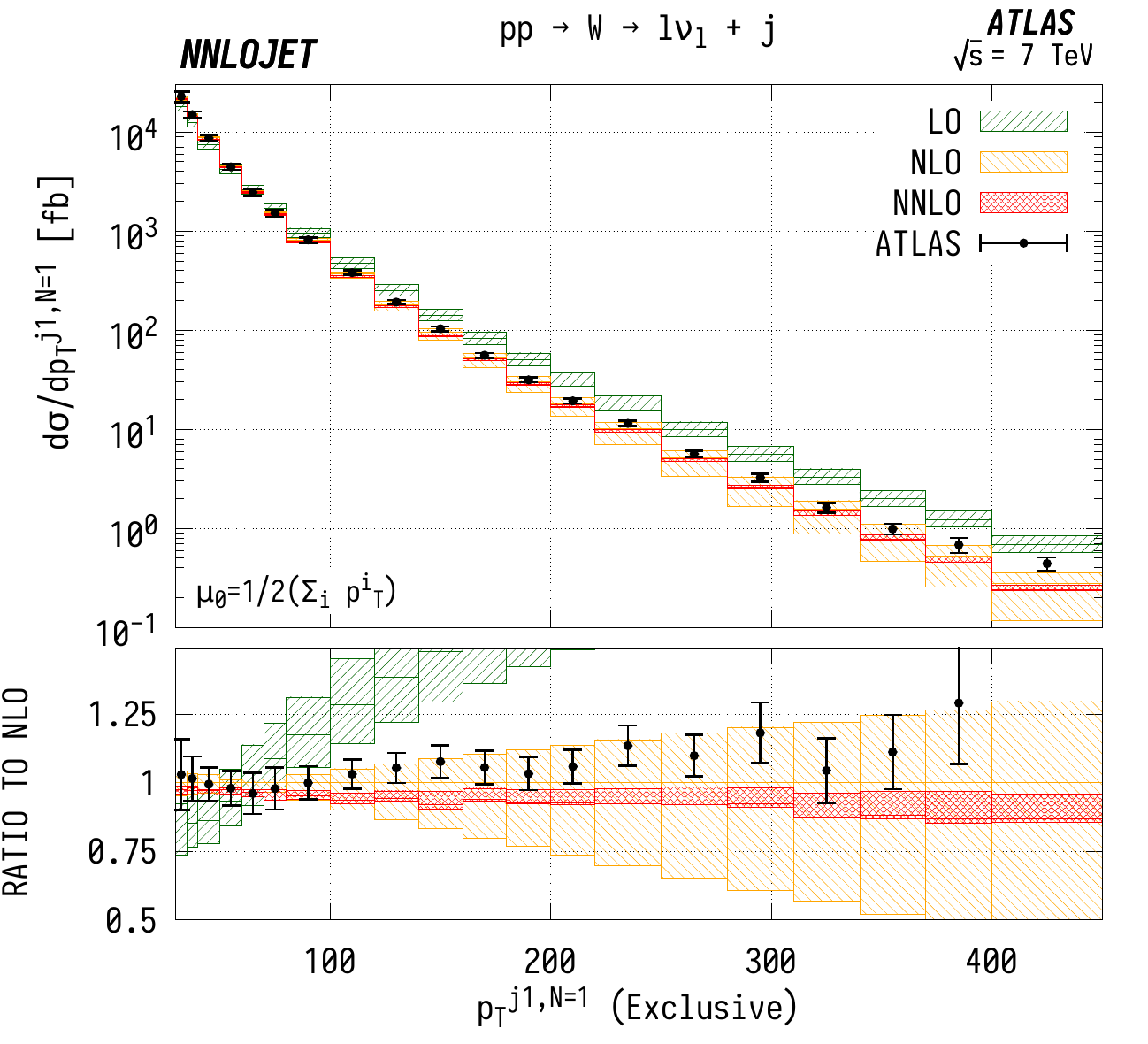}
  \caption{\label{fig:ATLAS_W_ptj1_excl}
    \PWJ cross section differential in the transverse momentum \pt of the leading jet for events with exactly one associated jet $(N=1)$ in the ATLAS fiducial region from Eq.~\ref{eqn:atlas_cuts}. Predictions at LO (green), NLO (orange), and NNLO (red) are compared to ATLAS data from Ref.~\cite{Aad:2014qxa}, and the ratio to NLO is shown in the lower panel. The bands correspond to scale uncertainties estimated as described in the main text.
  }
\end{figure}

\begin{figure}
  \centering
  \includegraphics[page=6,width=.95\linewidth]{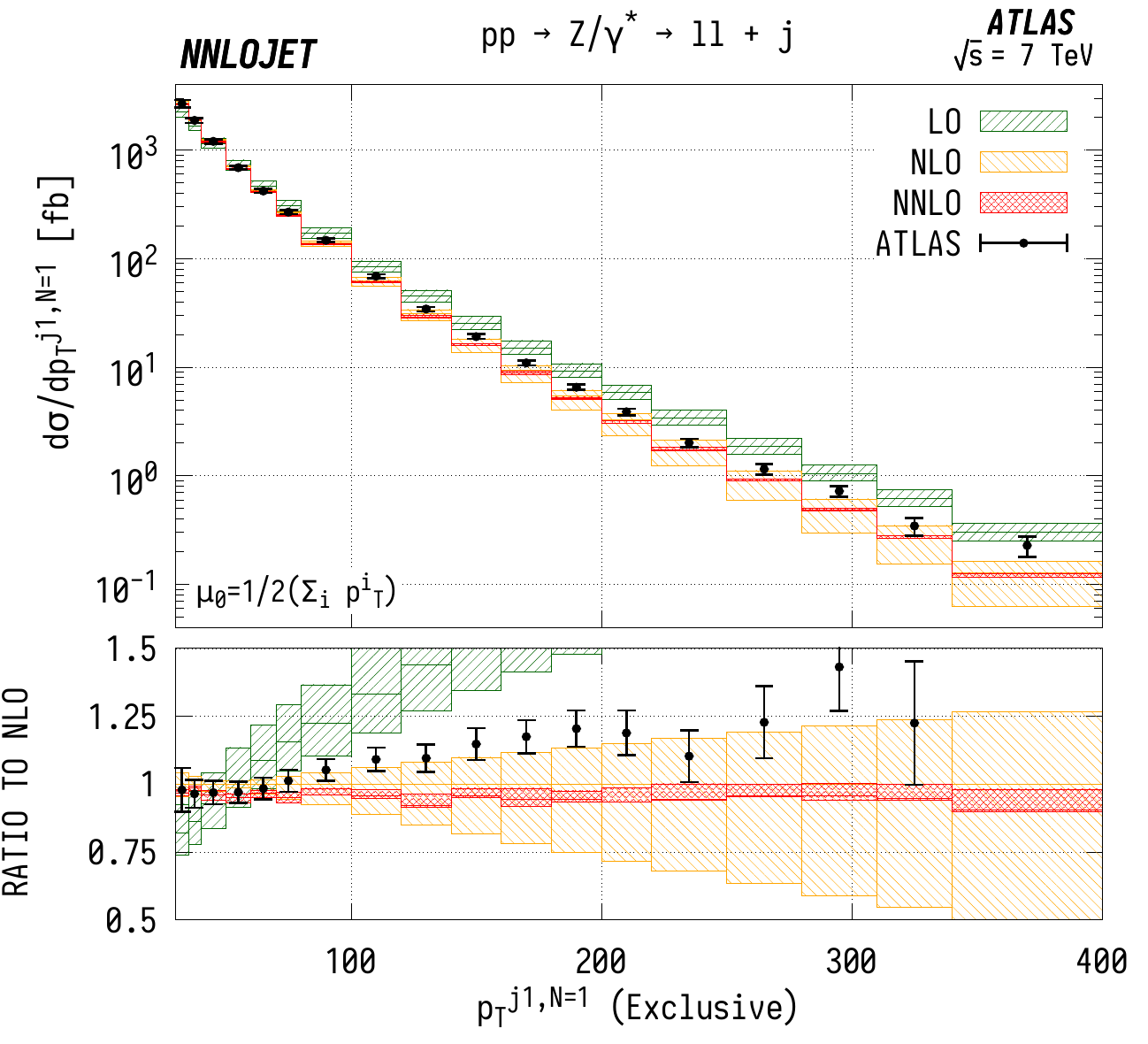}
  \caption{\label{fig:ATLAS_Z_ptj1_excl}
    \PZJ cross section differential in the transverse momentum \pt of the leading jet for events with exactly one associated jet $(N=1)$. Predictions at LO (green), NLO (orange), and NNLO (red) are compared to ATLAS data from Ref.~\cite{Aad:2013ysa}, and the ratio to NLO is shown in the lower panel. The bands correspond to scale uncertainties estimated as described in the main text.
  }
\end{figure}

First we consider the exclusive $(N=1)$ \pt~distribution of the leading jet for \PWJ~production using the data from~\cite{Aad:2014qxa} as shown in Fig.~\ref{fig:ATLAS_W_ptj1_excl}.
Here we observe agreement of the theory with data within errors up to $\ptjo \sim 80~\GeV$, beyond which the theoretical predictions are systematically below the data.
This behaviour is closely replicated in Fig.~\ref{fig:ATLAS_Z_ptj1_excl}, which shows the equivalent \PZJ~distribution. However beyond $\ptjo \sim 80~\GeV$,
the agreement with data is noticeably worse than for the \PWJ~distribution.
While we neglect electroweak corrections which have a well-known impact on the weak boson \pt~distributions~\cite{Denner:2009gj,Denner:2011vu,Kallweit:2015dum} from large Sudakov logarithms, these generally give considerable reductive K-factors at large \ptj and so would further worsen the agreement with data in both cases (see e.g.\ \cite{Kallweit:2015dum}).
For these exclusive distributions, it is instructive to note that \ptjo is equivalent to the transverse momentum of the vector boson due to the absence of extra jet radiation.

\subsection{Inclusive \ptjo Distributions}

\begin{figure}
  \centering
  \includegraphics[page=2,width=.95\linewidth]{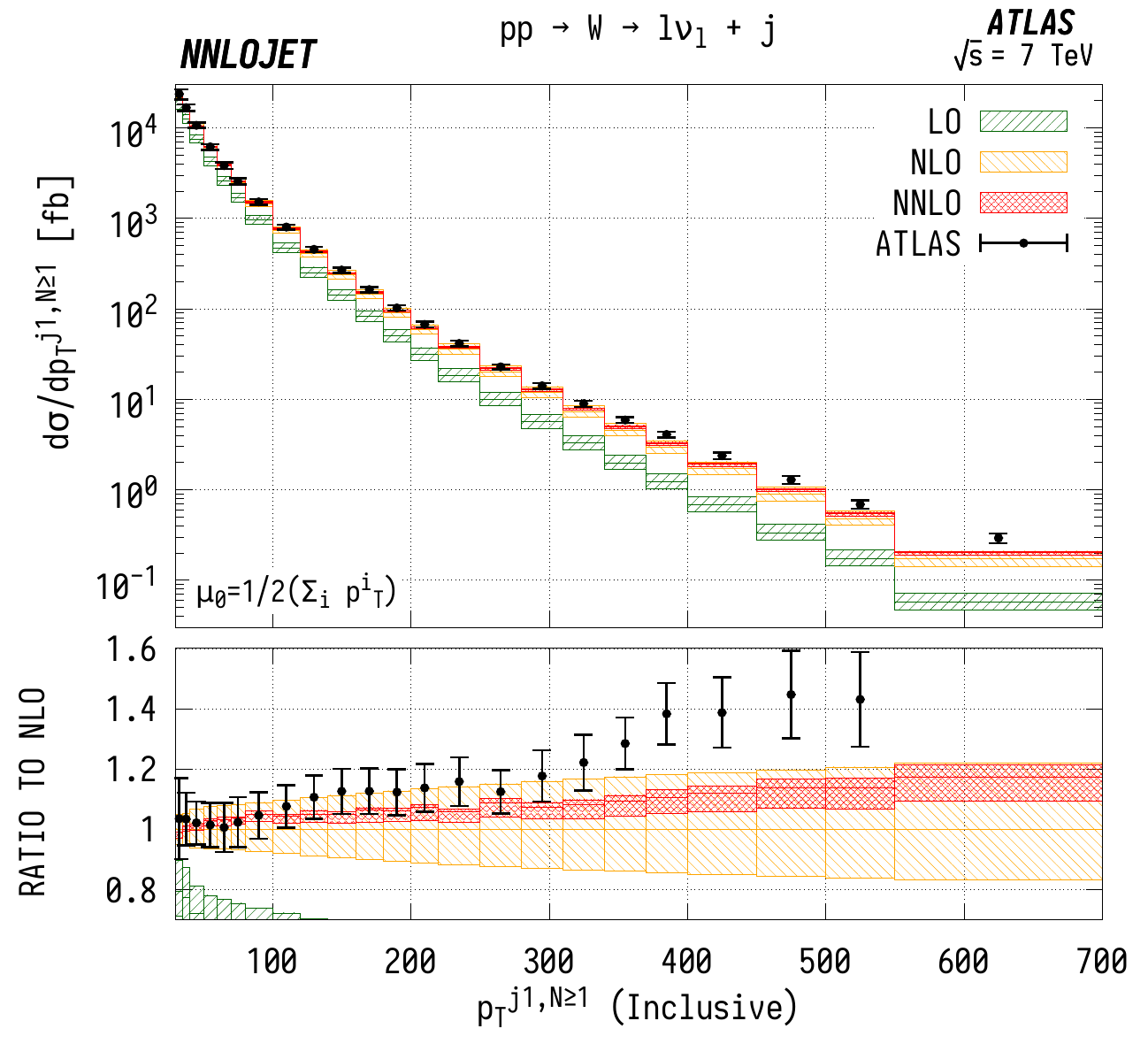}
  \caption{\label{fig:ATLAS_W_ptj1}
    \PWJ cross section differential in the transverse momentum \pt of the leading jet for events with one or more associated jets $(N\geq 1)$. See Fig.~\ref{fig:ATLAS_W_ptj1_excl} for details.
  }
\end{figure}
\begin{figure}
  \centering
  \includegraphics[page=5,width=.95\linewidth]{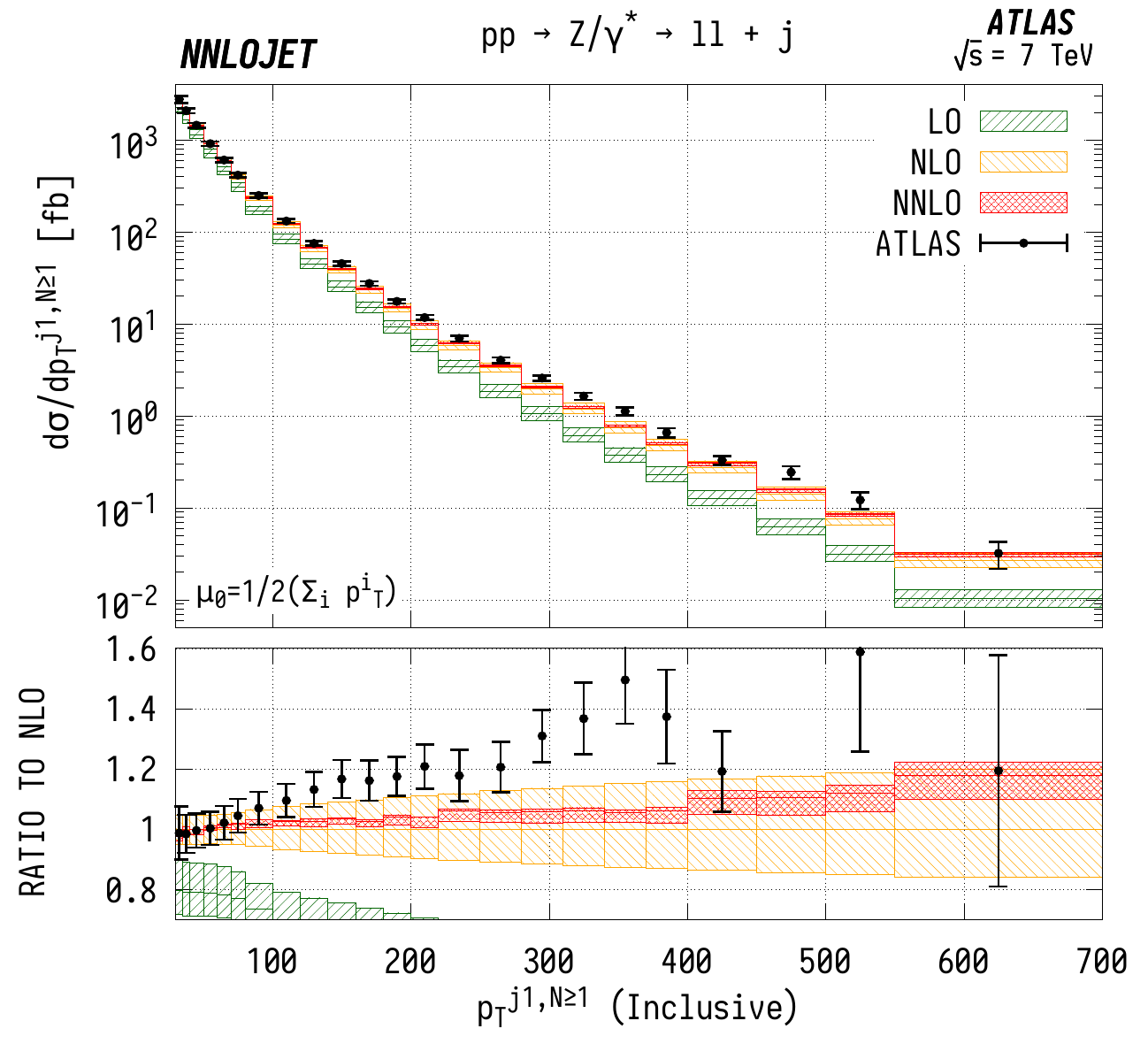}
  \caption{\label{fig:ATLAS_Z_ptj1}
    \PZJ cross section differential in the transverse momentum \pt of the leading jet for events with one or more associated jets $(N\geq 1)$. See Fig.~\ref{fig:ATLAS_Z_ptj1_excl} for details.
  }
\end{figure}

For the inclusive $(N\geq1)$ \ptjo~spectrum in \PWJ~production, shown in Fig.~\ref{fig:ATLAS_W_ptj1}, we observe marginally improved agreement over a wider range of \pt, with overlapping
uncertainty bands between data and theory up to $\ptjo \sim 300~\GeV$.
Beyond this point, there are substantial, $\cO(15\%)$, shape corrections when moving from NLO to NNLO which improve the agreement with data with respect to the NLO results.
In \PZJ~production, shown in Fig.~\ref{fig:ATLAS_Z_ptj1}, the pattern of perturbative corrections is very similar. However we do not observe the same
level of  improved agreement with data when moving from exclusive to inclusive jet production as for the \PWJ~process
and we again see that the theory prediction is systematically below the data from $\ptj\sim 100~\GeV$ onwards.

Allowing extra QCD radiation, as in the inclusive case, entails also allowing for dijet-type configurations where two hard jets are produced alongside a relatively soft vector boson.
In the full NNLO calculation, these $\cO(\alphas)$ contributions are first described at NLO, and give rise to a large QCD K-factors at high \ptj~\cite{Rubin:2010xp}.
This is the dominant cause of the distinct structure of the perturbative corrections between exclusive and inclusive production; for $N=1$ we see a decrease in the high-\ptjo~cross-sections with the inclusion of higher orders as opposed to an increase in $N\geq1$ production.
The difference in theory-to-data agreement between the \PZ~and \PW~distributions persists however, and may be a related to the different quark flavour combinations probed
by the different processes. Whilst not as constraining as the \PWp/\PWm~ratio, the \PW/\PZ~ratio still retains some sensitivity to the $u/d$ ratio due to different coupling strengths, and some dependence on the strange quark distributions, albeit suppressed compared to the inclusive Drell-Yan cross sections due to the Born-level gluon contribution.
The inclusion of higher-order EW terms are unlikely to describe the difference with respect to data at high \pt, as the EW corrections to the leading \ptj~distribution in vector-boson-plus-dijet events behave in a very similar manner for \PWJ~and \PZJ~production as demonstrated in~\cite{Kallweit:2015dum}.

\subsection{Exclusive/Inclusive Ratios}

\begin{figure}
  \centering
  \includegraphics[page=3,width=.95\linewidth]{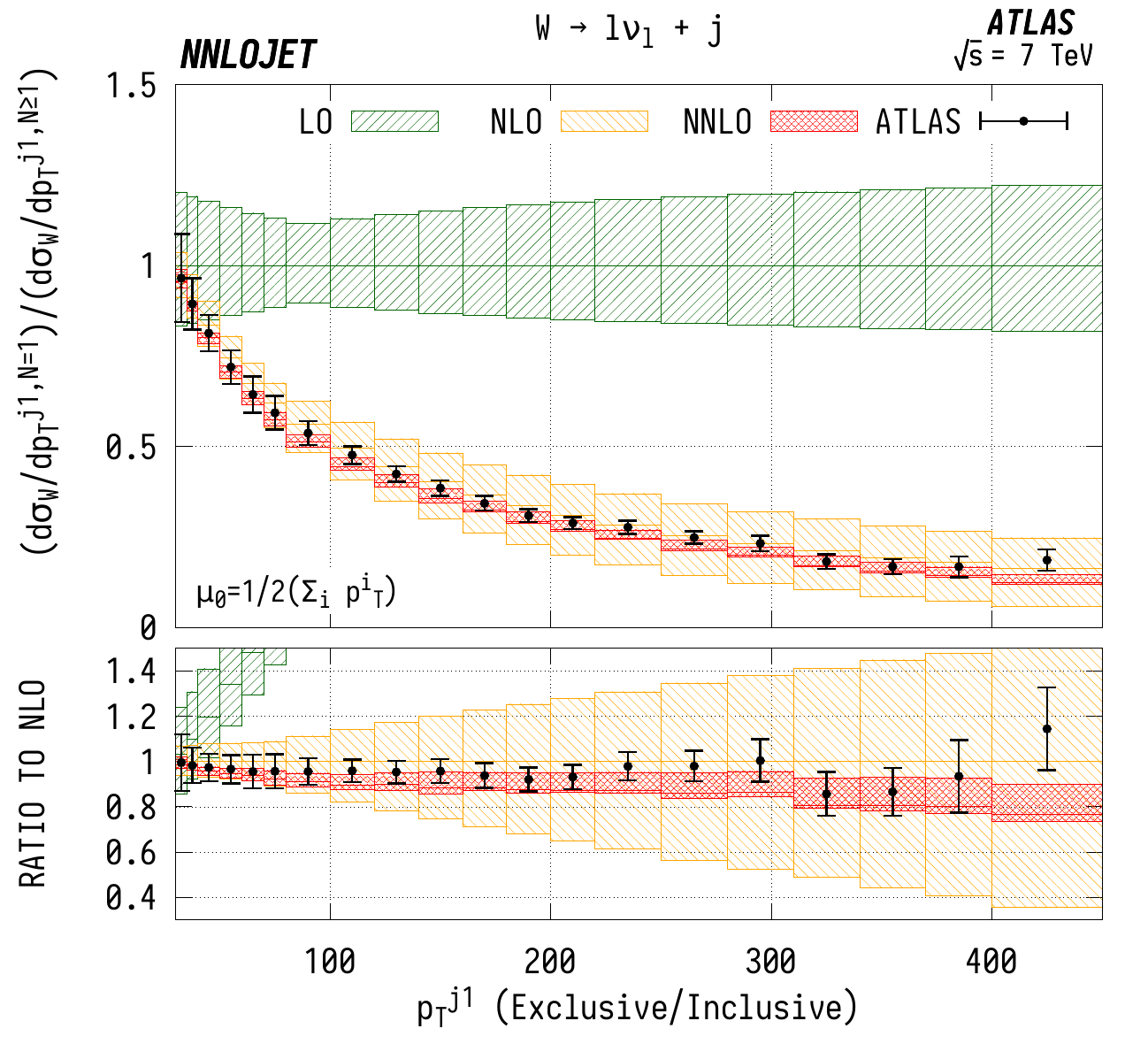}
  \caption{\label{fig:ATLAS_W_ptj1_exc_inc}
    Ratio of exclusive/inclusive $(N=1/N\geq 1)$ \PWJ~production differential in the transverse momentum \pt of the leading jet. Errors on the ATLAS data are approximated using uncertainties from the $N=1$ distribution normalised to the $N\geq 1$ results.  See Fig.~\ref{fig:ATLAS_W_ptj1_excl} for details.
  }
\end{figure}
\begin{figure}
  \centering
  \includegraphics[page=4,width=.95\linewidth]{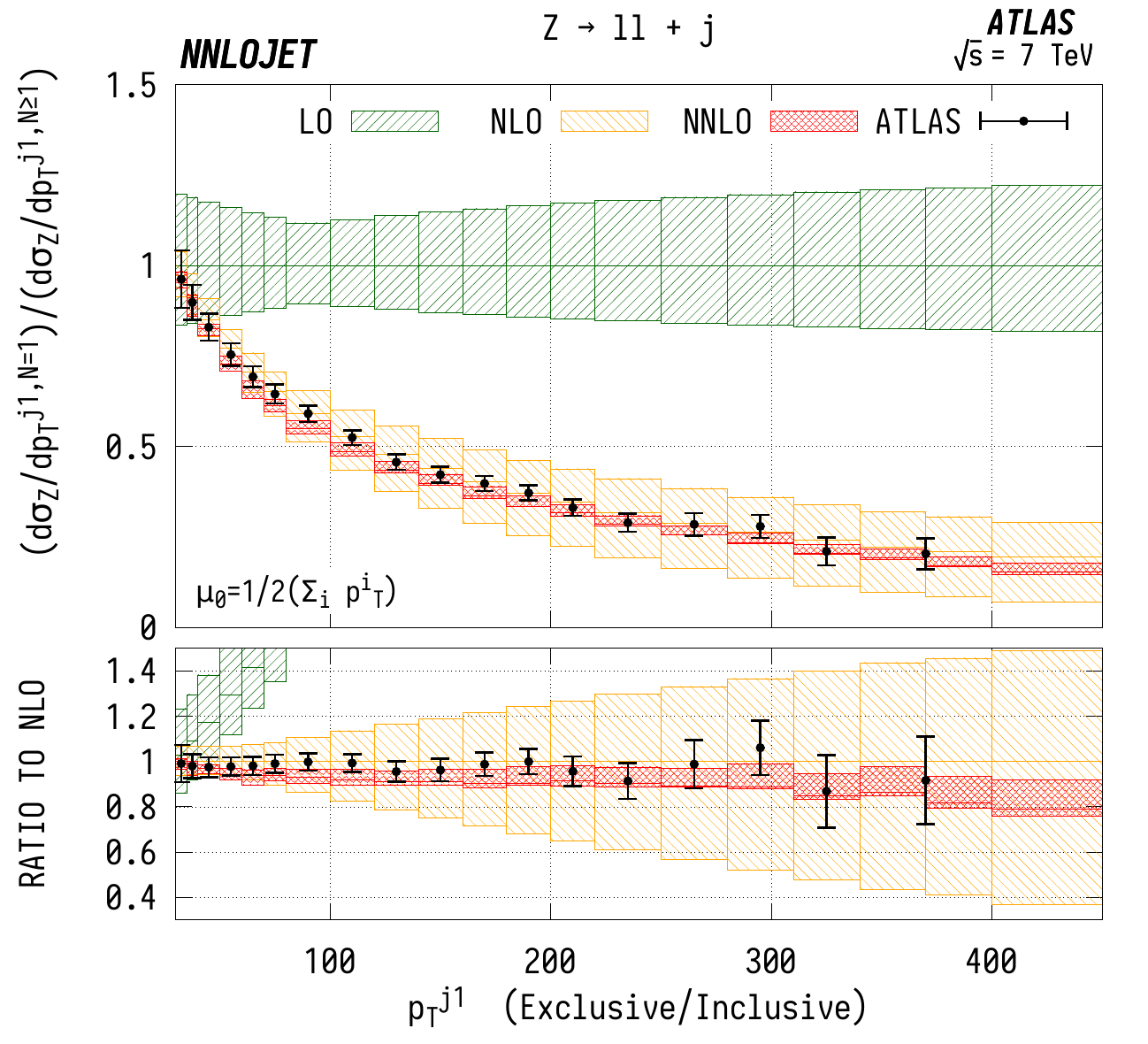}
  \caption{\label{fig:ATLAS_Z_ptj1_exc_inc}
     Ratio of exclusive/inclusive $(N=1/N\geq 1)$  \PZJ~production differential in the transverse momentum \pt of the leading jet. Errors on the ATLAS data are approximated using uncertainties from the $N=1$ distribution normalised to the $N\geq 1$ results.  See Fig.~\ref{fig:ATLAS_Z_ptj1_excl} for details.
  }
\end{figure}

In order to better understand the description of real emission  by the fixed order predictions, one can construct the ratio between the exclusive and inclusive leading jet distributions for both the \PWJ~and the \PZJ case,
shown in Figures~\ref{fig:ATLAS_W_ptj1_exc_inc} and~\ref{fig:ATLAS_Z_ptj1_exc_inc}.
The experimental measurements~\cite{Aad:2013ysa,Aad:2014qxa} do not explicitly quote the data in terms of exclusive/inclusive ratios.
We have therefore reconstructed it here using the central values of the relevant distributions with the errors approximated using uncertainties from the $N=1$ distribution normalised to the $N\geq 1$ results. For both distributions we observe similar behaviour, with good description of the data across the range of \ptjo, albeit with the general trend that the predictions systematically undershoot the central values of the data below $\ptjo\sim200~\GeV$, from which we can conclude that the extra jet radiation is well-described by the  fixed order predictions.


\subsection{W/Z Ratios Differential in Leading Jet \pt}

\begin{figure}
  \centering
  \includegraphics[page=14,width=.95\linewidth]{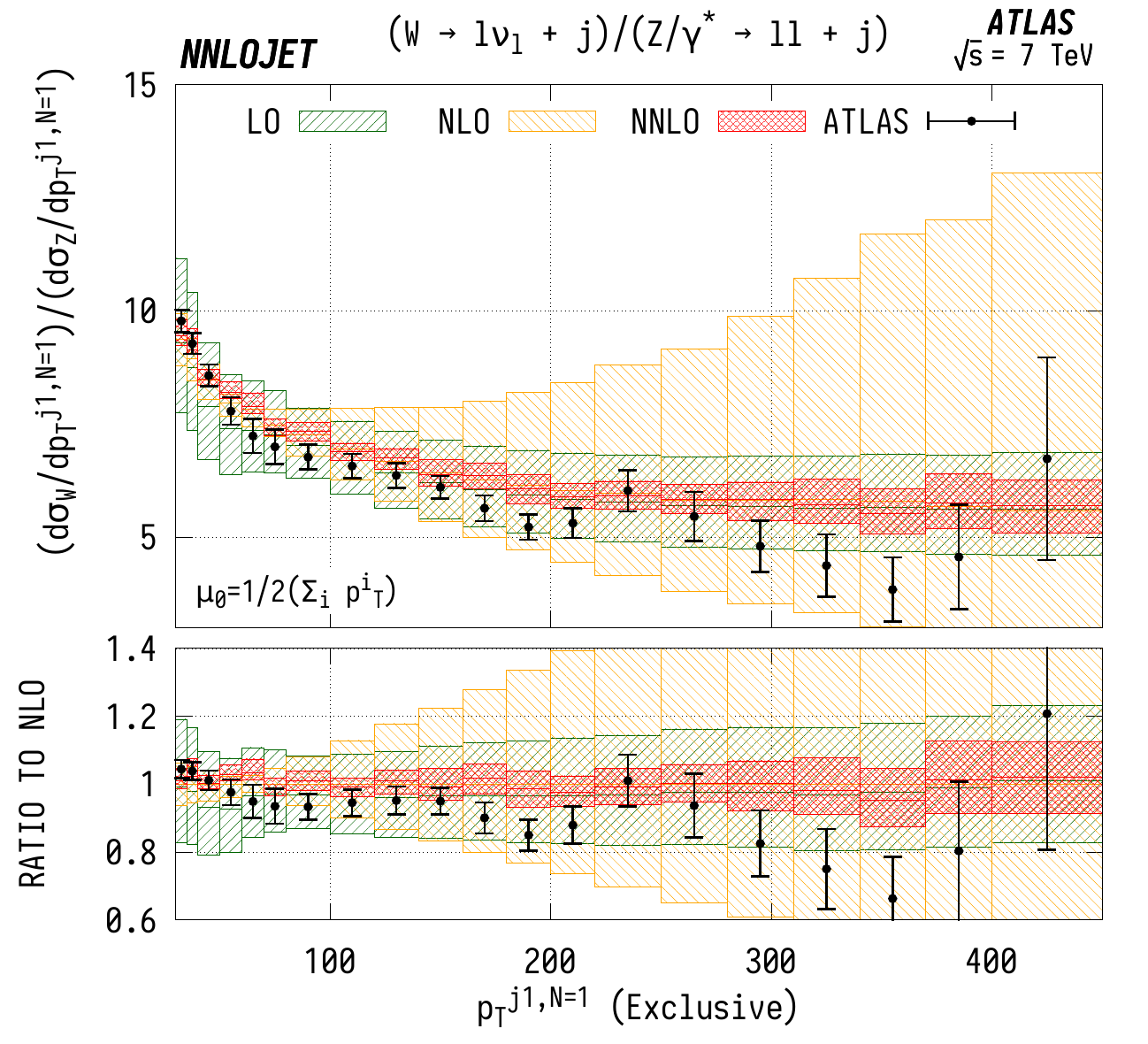}
  \caption{\label{fig:ATLAS_ptj1_exc}
    $\PWJ/\PZJ$ ratio differential in the exclusive \pt of the leading jet $(N=1)$. Predictions at LO (green), NLO (orange), and NNLO (red) are compared to ATLAS data from Ref.~\cite{Aad:2014rta}, and the ratio to NLO is shown in the lower panel. The bands correspond to scale uncertainties estimated as described in the main text.
  }
\end{figure}
\begin{figure}
  \centering
  \includegraphics[page=11,width=.95\linewidth]{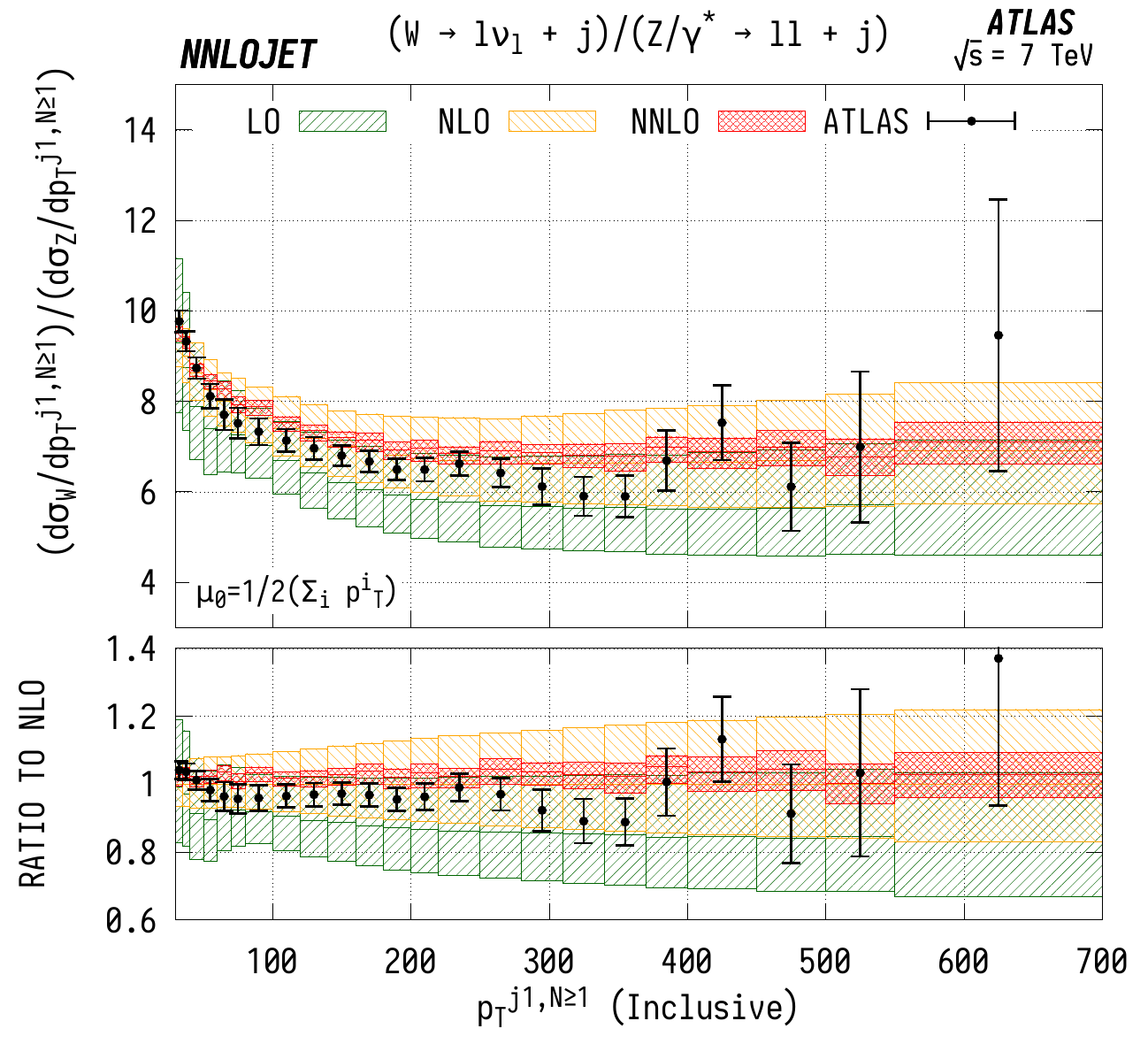}
  \caption{\label{fig:ATLAS_ptj1_inc}
    $\PWJ/\PZJ$ ratio differential in the inclusive \pt of the leading jet $(N\geq 1)$. See Fig.~\ref{fig:ATLAS_ptj1_exc} for details.
  }
\end{figure}

Figure~\ref{fig:ATLAS_ptj1_exc} shows the \PWJ/\PZJ ratio as a function of \ptjo,  for the exclusive $(N=1)$ case.
The large scale variation bands visible at NLO are a result of large NLO corrections at high \ptj that increase the scale uncertainties when propagated through ratios.
In particular, as shown in Fig.~\ref{fig:ATLAS_W_ptj1_excl} and Fig.~\ref{fig:ATLAS_Z_ptj1_excl},
we observe large reductive NLO/LO K-factors at high \ptjo for the individual \PZJ and \PWpmJ
distributions, reaching $K=0.3$ in the highest \ptjo bin, whereas the absolute size of the scale variation bands does not reduce significantly when going from LO to NLO.
This has the effect of making the exclusive $\PWJ/\PZJ$ ratio much more sensitive to scale variation in the constituent distributions at NLO than LO, artificially inflating the scale uncertainties at this order.
The inclusive ($N\geq 1$) ratio, shown in Fig.~\ref{fig:ATLAS_ptj1_inc}, has very similar central values at LO, NLO and NNLO, but does not display the inflated NLO scale uncertainty.

When taking the ratio, the impact of the extra jet activity is strongly suppressed, while the PDF sensitivity is enhanced.  As mentioned in the case of the individual
distributions, the \PW/\PZ~ratio can be used to provide constraints on the ratio of up and down valence quark distributions inside the PDFs, as well as on the strange distribution, due to the different couplings of the vector bosons. Taking only the dominant incoming $qg$ partonic configurations, we can see that na\"ively the ratio behaves as
\begin{equation}
  \frac{\sigma^\PWJ}{\sigma^\PZJ}\sim\frac{ug+dg}{0.29ug+0.37dg},
\end{equation}
where the numerical factors are the appropriate sums of the squares of the vector and axial vector quark-\PZ~couplings.
Discarding the common factor of the gluon PDF, this can be used to interpret a theory-to-data excess in the \PW/\PZ~ratio as an overestimate of the $u/d$ ratio.
If we look back to the individual distributions, we see that for each of the \PW~and \PZ~cases, the theory falls below the data. From this, it can be deduced that the most
probable cause is an underestimate in the $d$ quark content of the PDF.

\subsection{Inclusive Leading Jet Rapidity Distributions}

\begin{figure}
  \centering
  \includegraphics[page=1,width=.95\linewidth]{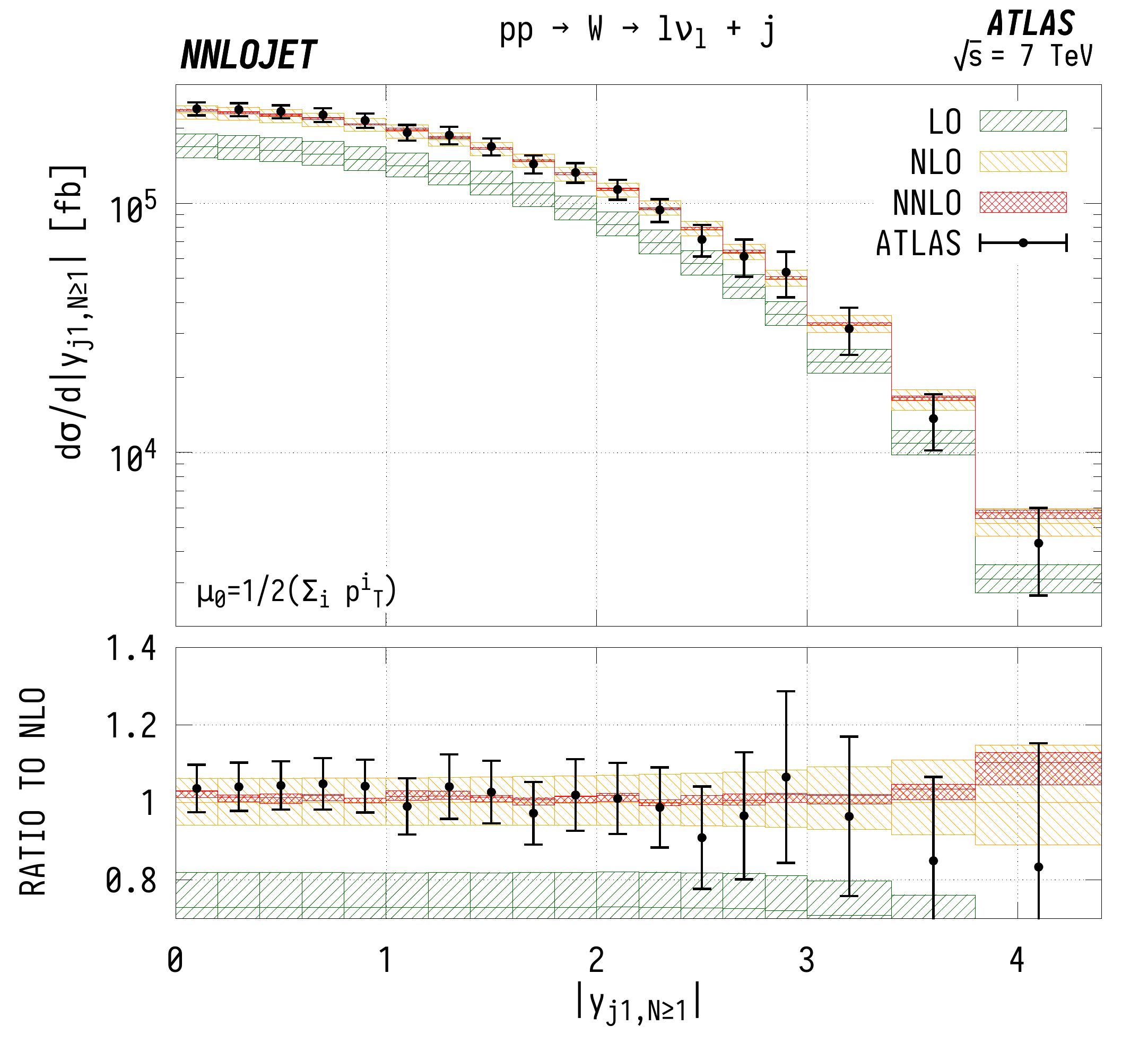}
  \caption{\label{fig:ATLAS_W_yj1}
    \PWJ cross section differential in the absolute rapidity $|y_j|$ of the leading jet.  See Fig.~\ref{fig:ATLAS_W_ptj1_excl} for details.
  }
\end{figure}

\begin{figure}
  \centering
  \includegraphics[page=4,width=.95\linewidth]{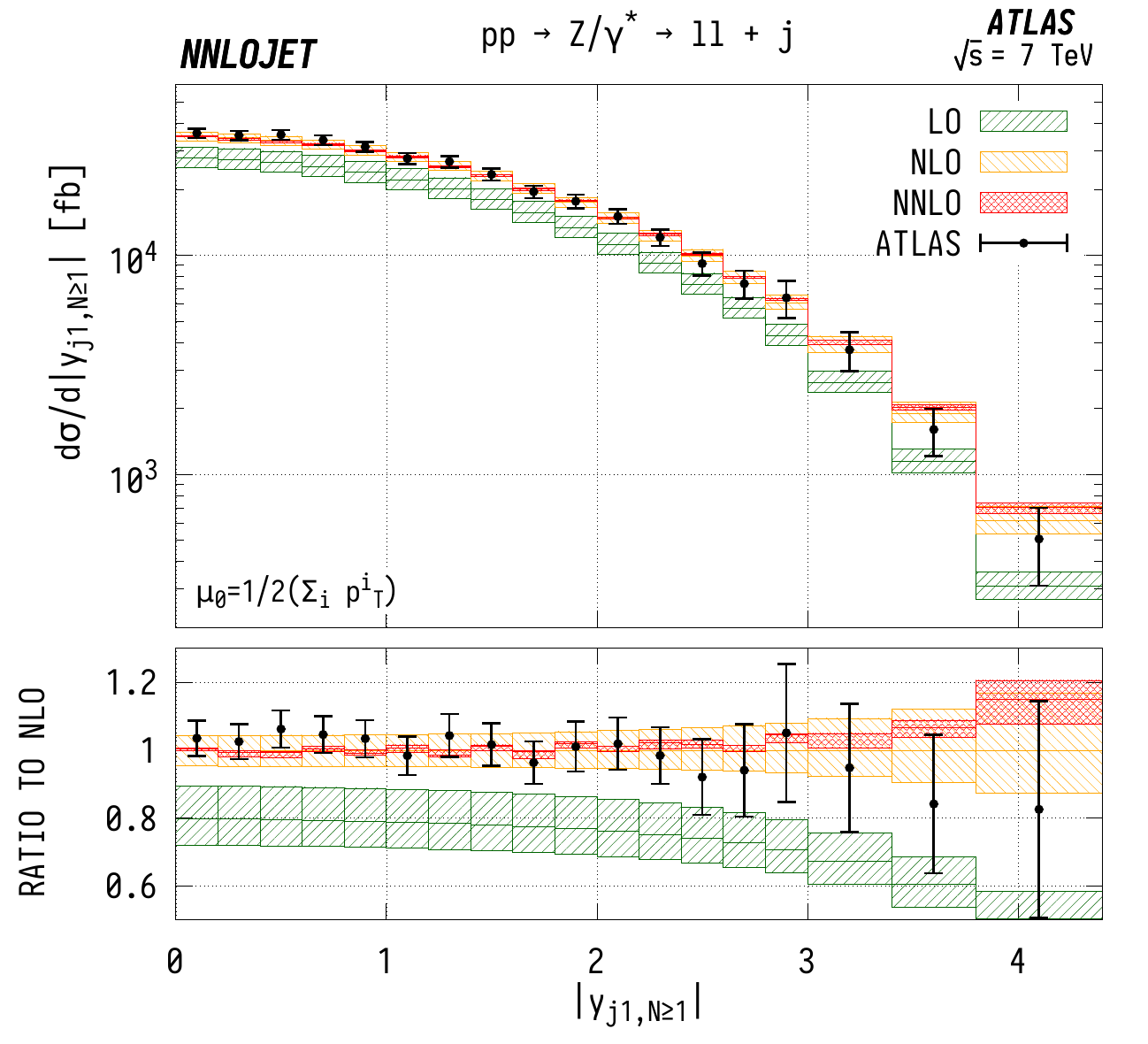}
  \caption{\label{fig:ATLAS_Z_yj1}
    \PZJ cross section differential in the absolute rapidity $|y_j|$ of the leading jet. See Fig.~\ref{fig:ATLAS_Z_ptj1_excl} for details.
  }
\end{figure}

The leading jet rapidity distribution $|y_{j1}|$ for \PWJ events is shown in Fig.~\ref{fig:ATLAS_W_yj1}, and for \PZJ events in Fig.~\ref{fig:ATLAS_Z_yj1}.
Here we observe that the higher-order QCD predictions are relatively stable for all orders up to $|y_{j1}| \sim 3$.
Beyond this point, we see a change in shape when transitioning from LO to NLO.
The shape is kept unmodified under the inclusion of the NNLO corrections.
There is an increase in scale uncertainty at higher rapidities $|y_{j1}|\gtrsim3.5$ due to large subleading jet contributions in this region, which are only described at lower orders for inclusive observables in the NNLO VJ calculation. In both cases, we see good agreement for all rapidities, with overlapping scale errors and experimental error bars for the entire distribution. However, the shape corrections induced at NNLO for $|y_{j1}|\gtrsim3.5$
modify the central values of the theory predictions such that the tension with data increases compared to NLO.


\begin{figure}
  \centering
  \includegraphics[page=8,width=.95\linewidth]{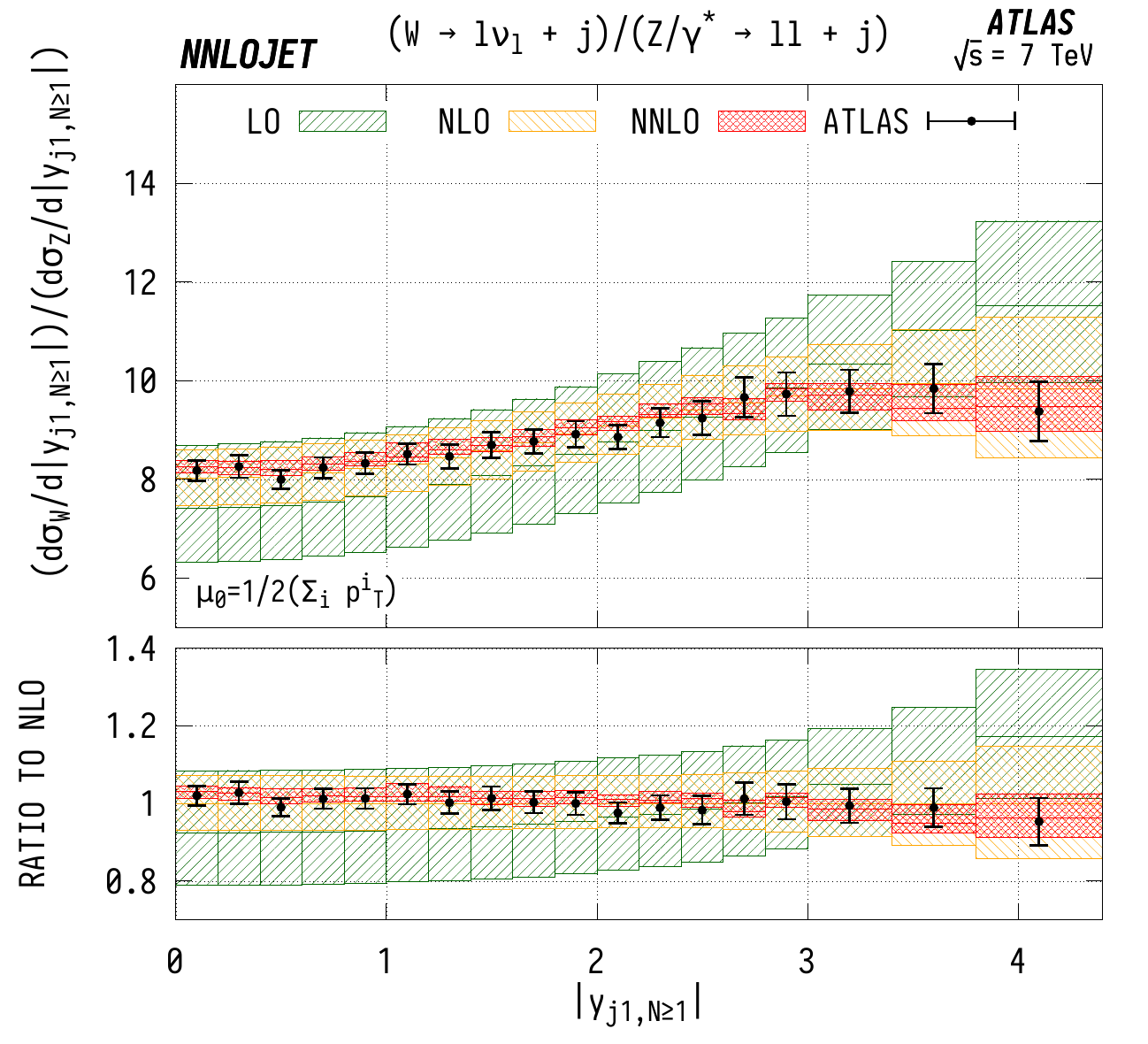}
  \caption{\label{fig:ATLAS_yj1}
    $\PWJ/\PZJ$ ratio differential in the absolute rapidity $|y_j|$ of the leading jet.  See Fig.~\ref{fig:ATLAS_ptj1_exc} for details.
  }
\end{figure}

If one associates the higher-energy incoming parton with $x_1$ and the lower-energy incoming parton with $x_2$, such that the sum of all final state momenta lies in the same direction as parton 1, the forward-most bin $(3.8<y_{j1}<4.4)$ in rapidity here corresponds to $x_1>0.19$, $x_2>0.00012$ for \PWJ production and $x_1>0.19$, $x_2>0.00019$ in \PZJ production.
One can then analyse the distributions here in a similar manner to the LHCb predictions in Figs.~\ref{fig:LHCb_Wm_etaj1}--\ref{fig:LHCb_Z_etaj1}. As is the case for the LHCb data, we see a
theory excess in the jet rapidity bins corresponding to $x\gtrsim 0.1$. This is again indicative of an overestimate of the gluon contributions to the PDF in this region since this excess is present in both \PW and \PZ distributions.
The central rapidity bins allow us to quantify better the PDF description at intermediate Bjorken-$x$, with the central-most bin in $y_{j1}$ requiring $x_1>0.0044$ and $x_2>0.0036$ for both neutral- and charged-current production.
Here we see good agreement with the data, indicating that the behaviour in this region is well under control. 

The ratio of \PWJ to \PZJ differential in the absolute rapidity $|y_{j1}|$ of the leading jet is shown in Fig.~\ref{fig:ATLAS_yj1}.
Due to the cross-cancellation in the ratios, we see that these
predictions display a considerably better perturbative stability than the individual distributions at high rapidities.
We observe excellent agreement with the ATLAS data across the entire rapidity range.
In the ratio, the PDF dependence of the predictions is in general lowered, particularly for gluonic contributions due to their similarity between the \PWJ and \PZJ cases.
The agreement on the ratio demonstrates that the NNLO QCD description of the underlying parton-level process is reliable.
It indicates that the discrepancies observed in the individual distributions are of parametric origin and can be remedied by an improved determination of the gluon distribution.

\section{Conclusions}


The recent calculations~\cite{Boughezal:2015dva,Ridder:2015dxa,Boughezal:2015ded,Boughezal:2016dtm,Boughezal:2016isb,Ridder:2016nkl,Gehrmann-DeRidder:2016jns,Campbell:2017dqk,Gauld:2017tww,Gehrmann-DeRidder:2017mvr} of NNLO QCD corrections to all observables related to the production of a massive vector boson in
association with a jet open up a new level of precision in the phenomenological interpretation of
these observables. In this context, final states at forward rapidity are particularly interesting, since they
correspond to initial states with very asymmetric momentum fractions of the incoming partons, thereby
probing the parton distributions in regions where they are insufficiently constrained by other data sets.

In this paper, we performed an in-depth comparison of forward
vector-boson-plus-jet data from LHCb~\cite{AbellanBeteta:2016ugk} and ATLAS~\cite{Aad:2014rta} with precise NNLO QCD
predictions, obtained using the \NNLOJET code~\cite{Ridder:2016nkl,Gehrmann-DeRidder:2016jns,Gehrmann-DeRidder:2017mvr}. Inclusion of NNLO QCD corrections leads to a substantial reduction of the theory uncertainty on the
predictions, thereby matching the accuracy of the LHC precision data.
Deviations between data and theory are observed in various distributions, which are further investigated
by constructing ratios between different vector bosons, and between inclusive and exclusive vector-boson-plus-jet
cross sections. The pattern of vector boson ratios and related asymmetries points to an overestimate of the
PDF parametrisation in the gluon distribution for Bjorken-$x\gtrsim 0.1$ and equally to an overestimate in
the $u/d$ quark ratio in the same region.


Our results highlight the unique sensitivity of forward vector-boson-plus-jet production to the PDF content of the proton.
We expect that the results presented here will enable improved determinations of the gluon distribution and of the quark flavour decomposition at large Bjorken-$x\gtrsim 0.1$, thereby enhancing the accuracy of theory predictions for
signal and background processes at the highest invariant masses.



\section{Acknowledgements}
The authors thank Rhorry Gauld for assistance with the uncertainty propagation in the LHCb data, and Xuan Chen, Juan Cruz-Martinez, James Currie, Marius H\"ofer, Imre Majer, Tom Morgan, Jan Niehues, Joao Pires and James Whitehead for useful discussions and their many contributions to the \NNLOJET code.
This research was supported in part by the UK Science and Technology Facilities Council, by the Swiss National Science Foundation (SNF) under contracts 200020-175595 and 200021-172478, by the ERC Consolidator Grant HICCUP (No. 614577) and by the Research Executive Agency (REA) of the European Union through the ERC Advanced Grant MC@NNLO (340983).

%
\end{document}